\documentclass[traditabstract]{aa} 
\usepackage{graphicx}
\usepackage{txfonts}
\usepackage{textcomp}
\usepackage{natbib}
\usepackage[breaklinks=false]{hyperref}
\usepackage{color}
\newcommand{\Ms}{{\ensuremath{{\rm M}_{\odot}}}}

\newcommand{\Ls}{{\ensuremath{{\rm L}_{\odot}}}}
\newcommand{\Teff}{{\ensuremath{T_{\rm eff}}}}
\newcommand{\logTeff}{{\ensuremath{\log(\Teff[\rm K])}}}
\newcommand{\Mpy}{\Ms\,{\rm yr}{\ensuremath{^{-1}}}}
\newcommand{\yr}{{\rm\ yr}}
\newcommand{\dm}{\ensuremath{\dot M}}
\newcommand{\gva}{{\sc genec}}
\newcommand{\syc}{{\sc syclist}}
\newcommand{\Mac}{\ensuremath{M_0}}
\newcommand{\Mz}{\ensuremath{M_{\rm ZAMS}}}
\newcommand{\hii}{H~{\sc ii}}
\newcommand{\rmand}{{\rm and}}

\begin{document}

\title{Stellar models and isochrones from low-mass to massive stars including pre-main sequence phase with accretion}
\titlerunning{Stellar models and isochrones including pre-main sequence phase with accretion}

\author{L. Haemmerl\'e, 
P. Eggenberger, 
S. Ekstr\"om, 
C. Georgy, 
G. Meynet, 
A. Postel, 
M. Audard, 
M. S\o rensen, 
T. Fragos} 
\authorrunning{Haemmerl\'e et al.}

\institute{D\'epartement d'Astronomie, Universit\'e de Gen\`eve, chemin des Maillettes 51, CH-1290 Versoix, Switzerland\label{inst1}}

\date{Received ; accepted }

 
 \abstract
{Grids of stellar models are useful tools to derive the properties of stellar clusters, in particular young clusters hosting massive stars,
and to provide information on the star formation process in various mass ranges.
Because of their short evolutionary timescale, massive stars end their life
while their low-mass siblings are still on the pre-main sequence (pre-MS) phase.
Thus the study of young clusters requires consistent consideration of all the phases of stellar evolution.
But despite the large number of grids that are available in the literature,
a grid accounting for the evolution from the pre-MS accretion phase to the post-MS phase in the whole stellar mass range is still lacking.
We build a grid of stellar models at solar metallicity with masses from 0.8 \Ms\ to 120 \Ms, including pre-MS phase with accretion.
We use the \gva\ code to run stellar models on this mass range.
The accretion law is chosen to match the observations of pre-MS objects on the Hertzsprung-Russell diagram.
We describe the evolutionary tracks and isochrones of our models.
The grid is connected to previous MS and post-MS grids computed with the same numerical method and physical assumptions,
which provides the widest grid in mass and age to date.
Numerical tables of our models and corresponding isochrones are available online.}
 

\maketitle
%

\section{Introduction}
\label{sec-in}

The star formation process, in particular that of massive stars, remains the least understood part of stellar evolution.
The complex physics involved (e.g. multi-dimensional hydrodynamics, turbulence, magnetic fields, radiative feedback)
makes the theoretical approach a challenging task.
On the other hand, direct observations of forming massive stars remain out of the reach of present-day telescopes
due to the optical thickness of their dust envelopes.
Given these limitations, indirect observations (e.g. observations of young clusters or young massive stars) coupled to theoretical models
have proven to be fruitful in providing constraints on the star formation process
(e.g.~\citealt{hayashi1961b,stahler1988,palla1990,palla1992,norberg2000,behrend2001,mottram2010,davies2011,haemmerle2016a}).

In recent years, the Geneva group has published a large number of stellar-evolution grids covering a wide mass- and metallicity range,
including the effect of rotation and mass loss
\citep{ekstroem2012,mowlavi2012,georgy2012,georgy2013a,georgy2013b,granada2014}.
These grids cover the whole MS phase and part of the post-MS phase.
Several grids including the pre-MS phase with accretion have also been computed (\citealt{bernasconi1996b,behrend2001}).
But since these works were published, the treatment of accretion in the \gva\ code has been significantly improved
(\citealt{haemmerle2014,haemmerle2016a,haemmerle2017}).
Moreover, despite the large variety of models that are currently available in the literature, a complete grid from low-mass to massive stars,
that covers all the phases of stellar evolution from the pre-MS accretion phase to more advanced stages of post-MS, is still lacking.

The inclusion of the pre-MS accretion is particularly important for massive stars which,
due to their short Kelvin-Helmholtz (KH) timescale, are thought to reach the MS before accretion is completed,
and their pre-MS evolutionary tracks reflect essentially their accretion history
\citep{bernasconi1996a,norberg2000,behrend2001,hosokawa2009,hosokawa2010,haemmerle2016a}.
On the other hand, for low-mass stars, whose KH time is several orders of magnitude longer than their accretion time
(e.g.~\citealt{larson1969,larson1972,stahler1983}), the pre-MS after the end of the accretion phase almost coincides
with a classical KH contraction at constant mass, and their evolution is well described by canonical non-accreting models.
The properties of accretion (rate, geometry, thermodynamics) have been found to impact the early constant-mass contraction of low-mass stars,
but this effect becomes small in a fraction of a KH time \citep{siess1997,baraffe2009,baraffe2012,hosokawa2011a,tognelli2015}.
Therefore, while accretion can be ignored in low-mass pre-MS grids, it is a required ingredient for grids that cover the early phases of massive star evolution.

Here, we present a new grid of stellar models at solar metallicity, covering nearly the whole present-day stellar mass range towards 120~\Ms,
including pre-MS with accretion.
We do not include rotation, and postpone it to a forthcoming publication.
The effect of binarity is also neglected (see \citealt{sorensen2018} for a recent study), as well as that of magnetic fields.

The paper is structured as follows.
In Sect.~\ref{sec-input}, we describe the assumptions used in the computations.
The models of the grid are described and discussed in Sects.~\ref{sec-grid} and \ref{sec-discus}, respectively.
We summarise our results in Sect.~\ref{sec-out}.

\section{Physical inputs to the models}
\label{sec-input}

\subsection{Stellar evolution code}
\label{sec-ge}

The models of the present grid have been computed with the 1D hydrostatic \gva\ stellar evolution code.
A detailed description of the code can be found in \cite{eggenberger2008}.
Here we recall the main ingredients: Convection is treated through the mixing-length theory, with the Schwarzschild criterion.
We use the same mixing-length parametre as \cite{ekstroem2012}:
for $\Mac<40$ \Ms, we take $\alpha=1.6$ where $\alpha$ is the ratio of the mixing length to the pressure-scale height;
for $\Mac>40$ \Ms, we take $\alpha=1.0$ where $\alpha$ is the ratio of the mixing length to the {\it density}-scale height.
We include overshooting, with the same prescriptions as \cite{ekstroem2012}.
We switch on overshooting only once accretion phase has ended.
Convective mixing is treated with a diffusive approach, since we expect the turnover time to be similar or longer than the D-burning timescale.
The chemical composition is solar ($X=0.72$, $Y=0.266$ and $Z=0.014$), in the initial low-mass seed (homogeneous)
as well as in the accreted material.
We use the abundances of \cite{asplund2005} with the neon mass fraction of \cite{cunha2006},
and include deuterium and lithium with mass fractions $X_2=5\times10^{-5}$, $X_6=9\times10^{-10}$ and $X_7=10^{-8}$
\citep{bernasconi1996a,bernasconi1996b,norberg2000,behrend2001,haemmerle2013,haemmerle2016a}. Energy production includes gravitational contraction and nuclear reactions, with the rates of \cite{caughlan1988}.
We include all the reaction networks used in \cite{ekstroem2012}, with in addition deuterium- and lithium-burning.
Atomic diffusion due to concentration and thermal gradient is included only for models with a mass on the ZAMS lower than 1.3~\Ms,
and only once accretion is completed.
Mass-loss is also included, with the same prescriptions as \cite{ekstroem2012}.
However, we do not treat mass-loss and accretion simultaneously.
Mass-loss is switched on only at the end of the accretion phase.

\subsection{Accretion history}
\label{sec-dm}

Accretion history in massive star formation is an open question.
The complex physics of the pre-stellar collapse (e.g. turbulence, magnetic fields, radiative feedback)
prevents firm conclusions from being drawn on the detailed behaviour of the mass accretion rate \dm.
A large number of works have studied the collapse of massive pre-stellar clouds from the theoretical point of view,
using various numerical approaches and physical inputs
(e.g.~\citealt{yorke1977,yorke2002,mckee2003,schmeja2004,krumholz2007,krumholz2009,peters2010a,peters2010b,peters2010c,peters2011,
kuiper2010b,kuiper2011, girichidis2011,girichidis2012a,girichidis2012b,meyer2017,meyer2018}).
Beyond the general agreement on typical rates of $\sim10^{-3}$ \Mpy\ for massive stars, the main features of accretion histories differ significantly between various studies.
Some models show an early peak in \dm\ followed by a slow decrease \citep{yorke1977,yorke2002,schmeja2004,kuiper2010b,kuiper2011},
some show a continuously growing rate \citep{mckee2003,girichidis2011,girichidis2012a,girichidis2012b,meyer2017,meyer2018},
and some show a more complex, highly stochastic history \citep{krumholz2007,krumholz2009,peters2010a,peters2010b,peters2010c,peters2011}.

\begin{figure}
\includegraphics[width=0.49\textwidth]{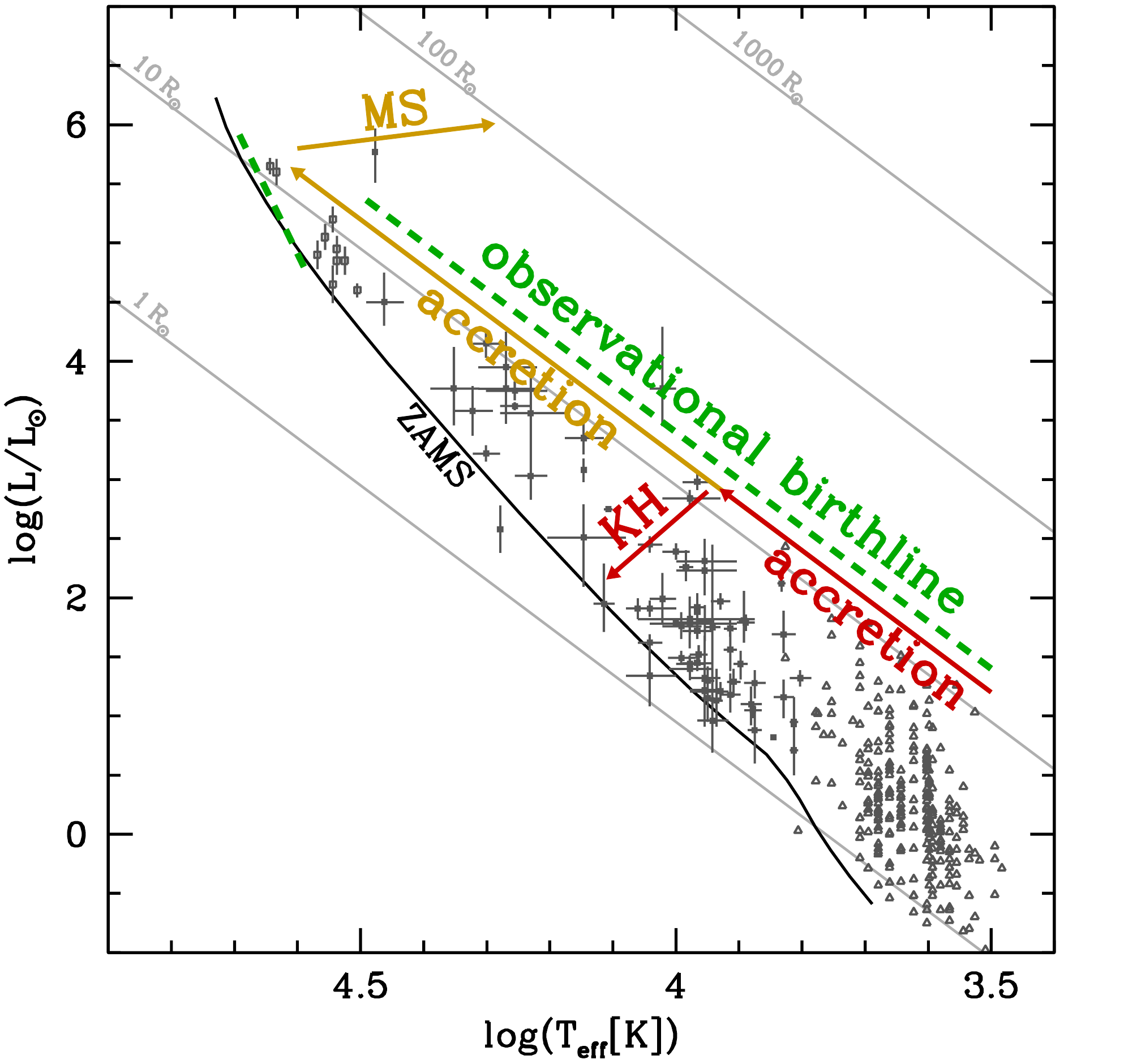}
\caption{Schematic diagram of the birthline.
The dots are observations of T Tauri stars (\citealt{cohen1979}, triangle),
Herbig Ae/Be stars (\citealt{alecian2013a}, solid squares), and young O stars (\citealt{martins2012}, open squares).
Error bars are indicated when available.
The black solid line is the ZAMS of \cite{ekstroem2012}, and the grey straight lines are iso-radius of indicated radii.
The observational birthline (upper envelope of the observation) is illustrated in green.
Red and orange arrows represent schematic tracks of intermediate-mass (red) and massive (orange) stars.}
\label{fig-blobs}
\end{figure}

An alternative way to establish the accretion history of massive stars is to use observations.
Directly observing the accretion process on a forming massive star is far beyond the reach of present-day telescopes,
but indirect considerations can provide constraints on the way massive stars grow in mass.
For instance, it is possible to compare evolutionary tracks of accreting stars with sets of observed pre-MS objects.
This method has been widely used in the literature (e.g.~\citealt{stahler1988,palla1990,palla1992,norberg2000,behrend2001}).
This method relies on the fact that stars become optically visible when they stop accreting,
because the end of accretion is assumed to correspond to the dispersal of their surrounding gas and dust.
This is of course a simplification, since accretion could proceed through a disc even after the dispersal of the outer envelope,
which could make the star visible before the end of accretion.
Here, we assume that such a configuration occurs only after the end of the main accretion phase.
Observationally, the birthline is defined as the line on the Hertzsprung-Russell (HR) diagram
along which young stars become visible in the optical (Fig.~\ref{fig-blobs}).
Since pre-MS stars are contracting, the birthline corresponds to the upper (high-$L$) envelope of observed pre-MS stars.
It joins the ZAMS at $10^{5-6}$ \Ls, as show the observations in Fig.~(\ref{fig-blobs}),
which is understood by the short KH time of pre-MS stars with higher luminosities.
These stars achieve their contraction before finishing accretion,
meaning that they are already on the MS when they emerge from their natal cloud.
This implies that the birthline of massive stars corresponds to the {lower} envelope of the observations
(illustrated by the short green dashed line on Fig.~\ref{fig-blobs}, along the upper end of the ZAMS).

\cite{churchwell1999} and \cite{henning2000} established an empirical correlation between
the mass outflows $\dm_{\rm out}$ through jets in ultra-compact \hii\ regions and the bolometric luminosity $L$ of their central source,
the Churchwell-Henning (CH) relation.
A polynomial fit gives (\citealt{behrend2001})
\begin{eqnarray}
\log\,(\dm_{\rm out})=-5.28+0.752\,\log{L\over\Ls}-0.0278\,\log^2{L\over\Ls}
\label{eq-chout},\end{eqnarray}
where $\dm_{\rm out}$ is in \Mpy.
Thus, the more luminous the protostar, the stronger the mass outflows.
This correlation can be interpreted in two different ways: as an evolutionary sequence
or as a spread in the `final' mass \Mac\ of the accreting protostar\footnote{\ 
   In the context of accretion, the final mass of the star is the mass at the end of accretion.
   This is in contrast with MS models with mass loss, where this mass is called `initial'.
   In the present work, we consider both accretion and mass loss.
   In order to avoid ambiguity, we use \Mac\ to denote the mass at the end of accretion, before any mass loss.
    \Mac \ corresponds to the maximum mass of the model, once accretion is achieved and mass loss starts.\label{not-m0}}.
In the first interpretation, as an accreting protostar grows in mass, its luminosity and outflows increase together according to Eq.~(\ref{eq-chout}).
In the second one, protostars destined to reach various masses drive mass outflows that reflect their luminosity at the end of the accretion phase,
that is,  they reflect their `final' mass \Mac.
In this case, the relevant value of $L$ in Eq.~(\ref{eq-chout}) is the luminosity corresponding to the `final' mass \Mac, not to the current mass $M(t)$.
In the present work, we follow the first interpretation.
The impact of this assumption is discussed in Sect.~\ref{sec-ac2}.

Assuming the mass accretion rate onto the central source is proportional to the mass outflow,
\cite{behrend2001} and \cite{haemmerle2013,haemmerle2016a} derived accretion histories
and applied them to pre-MS models for massive star formation.
From the mass inflow at the inner edge of the disc, $\dot M_{\rm disc}$, a fraction $f$ is assumed to be effectively accreted by the star,
the rest being ejected through the bipolar outflows.
This gives the accretion rate onto the star as a function of $\dot M_{\rm out}$,
\begin{equation}
\dot M=f\,\dot M_{\rm disc}={f\over1-f}\,\dot M_{\rm out},
\label{eq-chf}\end{equation}
that is, as a function of the bolometric luminosity $L$, through Eq.~(\ref{eq-chout}).
The value of $f$ is then determined by comparing the evolutionary track of accreting stars with observations of pre-MS stars.

The accreting models of \cite{behrend2001} and \cite{haemmerle2013,haemmerle2016a}
are based on the assumption of a unique accretion history for stars of various \Mac;
this reflects the choice of the interpretation of the CH relation as an evolutionary sequence, since $f$ is assumed to be unique for a given mass.
This also means that all stars accrete according to the same $M-\dm$ relation,
and that their accretion rate depends only on their current properties (age, mass, and luminosity).
This assumption can be partly justified by Newton's theorem.
This unique accretion rate is given by the CH relation, that is, Eqs.~(\ref{eq-chout}) and (\ref{eq-chf}).
In this scenario, each individual star follows the same evolutionary track while it accretes mass.
The total mass \Mac\ accreted by the star during the whole accretion phase depends only on the age at which accretion stops.
At this point, the track switches towards that of constant mass evolution and the star becomes optically visible.
Thus this unique accretion track must correspond to the observational birthline (Fig.~\ref{fig-blobs}).
By convention, this accretion track is called the (theoretical) birthline.
If accretion stops before the star has reached the ZAMS, it follows a KH contraction towards the ZAMS (red arrows on Fig.~\ref{fig-blobs}).
If accretion stops only after the star reaches the ZAMS, it switches directly to MS evolution (orange arrows on Fig.~\ref{fig-blobs}).

These considerations allowed \cite{behrend2001} and \cite{haemmerle2016a} to estimate the value of $f$.
\cite{haemmerle2016a} showed that the best fit was obtained with a value of $f$ that decreases as the stellar mass increases,
from $f=1/3$ in the low-mass range to $f=1/11$ in the high-mass range.
This unique accretion history corresponds to an accretion rate that increases strongly with the current stellar mass,
from $\sim10^{-5}$~\Mpy\ in the low-mass range to $\sim10^{-3}$~\Mpy\ in the high-mass range.

This picture of an accretion rate that increases with time was recently confirmed by \cite{yang2017},
through the analysis of clumps from the ATLASGAL survey.
We also note that \cite{davies2011} studied the distribution in luminosity of massive young stellar objects in the Milky Way
and compared it with the predictions of various models.
These latter authors obtained a better agreement using accretion histories with rates increasing with time than with constant or decreasing rates.
Moreover, a unique accretion history is able to reproduce the observations.
Finally, an accretion rate that increases with time results also from the simplest theory of accretion,
the Bondi-Hoyle mechanism \citep{bondi1944,bondi1952}, where $\dm\propto M^2$.

\begin{figure}
\includegraphics[width=0.49\textwidth]{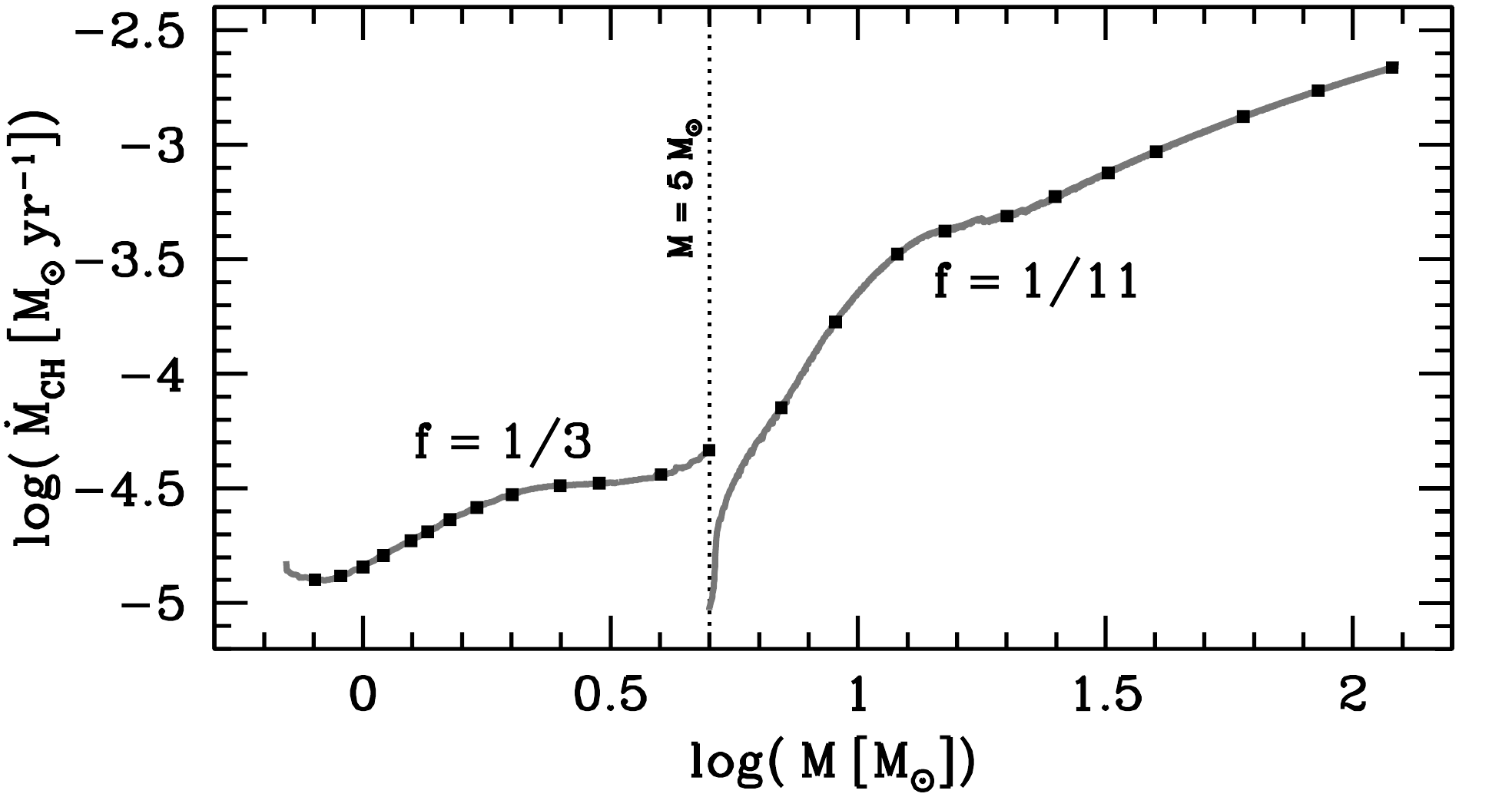}
\caption{Churchwell-Henning accretion rate as a function of the stellar mass.
The vertical dotted line indicates $M=5M_\odot$, the mass at which the value of $f$ is switched from 1/3 to 1/11.
The black squares indicate the masses used to build the stellar grids.}
\label{fig-dm}
\end{figure}

In the present work, we use the CH accretion rate with a value of $f$ switched at $M=5\,M_\odot$, according to \cite{haemmerle2016a}:
\begin{equation}
f=\left\{\begin{array}{ll}
1/3     \quad &M\leq5M_\odot    \\
1/11    \quad &M>5M_\odot               \end{array}\right..
\label{eq-f}\end{equation}
This switch in $f$ introduces a discontinuity in the accretion rate, but not in the age-mass relation.
By computing a birthline with this prescription, the stellar models provide the relation between the current stellar mass and the stellar luminosity,
and thus a relation between the current stellar mass and the accretion rate.
This $M-\dot M$ relation is shown in Fig.~\ref{fig-dm}.
We emphasise that this relation is model dependent, since it relies on the mass--luminosity relation given by the stellar models.

As mentioned above, the CH relation does not correspond necessarily to an evolutionary sequence,
and might reflect instead the properties of stars during their main accretion phase.
In this alternative interpretation, protostars destined to become massive would start their accretion phase
already at the high rates expected for massive star formation ($\sim10^{-3}$ \Mpy), even if their current mass is still low.
This would suggest that it is more pertinent to interpret the $M-\dot M$ relation of Fig.~\ref{fig-dm} with $M=\Mac$ instead of $M=M(t)$.
Although we do not expect these two interpretations to lead to significant differences in terms of birthline and evolutionary tracks,
the impact on the isochrones could be significant.
Indeed, in the context of a unique accretion history given by the CH relation, massive stars spend a long time in the low-mass range,
accreting at the corresponding low rate (see Fig.~\ref{fig-dm}).
In contrast, if stars that are destined to reach high masses start their accretion phase
already with the high rates expected for massive star formation, they will spend a much shorter time as low-mass objects,
and reach the location of massive stars on the HR diagram at a younger age.
This effect is discussed in Sect.~\ref{sec-ac2}.

\subsection{Thermal properties of the accreted material}
\label{sec-th}

Once the accreted material is incorporated in the stellar interior, we have to define its thermal properties, essentially its entropy content.
In all the models described in the present work, we use the so-called {cold disc accretion} (\citealt{palla1992,hosokawa2010}).
This assumption is the same as that used in all the previous publications of the Geneva group that included accretion
(e.g.~\citealt{bernasconi1996a,norberg2000,behrend2001,haemmerle2013,haemmerle2016a,haemmerle2017}).
It states that the thermal properties of the accreted material match those of the stellar surface before the material is accreted.
In other words, the internal profiles of all the thermal quantities, in particular the entropy, are built continuously.
This assumption corresponds to the case of accretion through a geometrically thin disc.
The physical justification is that if the disc is thin enough any entropy excess is radiated away in the polar directions
before being advected in the stellar interior.

Cold disc accretion is the lower limit for the accretion of entropy,
the upper limit being spherical accretion (e.g.~\citealt{hosokawa2009}), where all the entropy of the accretion flow is advected inside the star.
Using the homology relations, one has
\begin{equation}
s\sim\ln{T^{3/2}\over\rho}\sim\ln{MR^3},
\end{equation}
that is, at a given mass, the higher the entropy content of a star, the larger its radius.
As a consequence, using the assumption of spherical accretion would lead to larger radii than those reached in our models,
as seen in \cite{hosokawa2009}.
However, it is expected that such a spherical geometry is only relevant during the early accretion phase:
in most of the accretion phase followed in our models, accretion is expected to proceed through a disc.
In fact, since the CH relation that provides the accretion rates is based on observations of bipolar outflows,
it assumes implicitly a disc-like accretion geometry during the main accretion phase.

We note that the assumption of cold accretion is expected to be relevant only for rates $\lesssim10^{-4}$ \Mpy\ \citep{baraffe2012}.
For higher rates, the advection of entropy is too efficient for the excesses to be radiated away, even when accretion proceeds through a disc.
Nevertheless, our accretion rates only exceed $10^{-4}$ \Mpy\  at the end of the swelling
triggered by the transition from convective to radiative structures.
After this stage, the choice of the entropy accreted is not thought to significantly impact the structure \citep{hosokawa2010}.
Thus cold accretion is relevant for the present case, thanks to the choice of the accretion rate.

\subsection{Initial model}
\label{sec-ini}

In all the models of the present work, we use the same initial conditions, given by
\begin{equation}\begin{array}{ll}
M=0.7\,\Ms\ ,   \qquad& Z=0.014\ ,      \\
L=9.57\,\Ls\ ,  \qquad\rmand\quad&      \Teff=4130\rm\,K.
\end{array}\end{equation}
This initial model has a radius of $6.06\,R_\odot$ and is fully convective.
To avoid problems in numerical convergence of the models at low masses, we decided to skip the early accretion phase,
which should begin with a seed mass of the order of magnitude of about 0.01 \Ms\ (i.e. the second Larson core),
and start our computation from a larger mass model on the birthline as that used in \cite{haemmerle2016a}.

\cite{haemmerle2016a} and \cite{haemmerle2016b} showed that for rates $10^{-4}-10^{-3}$~\Mpy,
the choice of the initial model has a significant impact on the evolutionary track during the accretion phase,
more precisely during the swelling produced by the internal luminosity wave (\citealt{larson1972,hosokawa2009,hosokawa2010}).
But it was found that for the CH accretion rate, this effect is negligible,
since the accretion rate is still $\sim10^{-5}-10^{-4}$~\Mpy\ when the star enters the swelling phase.
Therefore, initial conditions do not significantly impact the evolution in the present context.
This could change if we relax the assumption of cold accretion.

Since we do not start the run at $M=0$, in order to avoid an artificial shift in age, we consider a non-zero initial age.
We assume instead that the 0.7~\Ms\ seed formed by accretion from $M=0$ at a constant rate, given by the CH rate for $M=0.7$~\Ms,
which is $1.157\times10^{-5}\,\Mpy$.
Thus the age of the initial model is defined by
\begin{equation}
t_{\rm ini}={0.7\,\Ms\over1.157\times10^{-5}\,\Mpy}=60\ 500\yr.
\label{eq-t0}\end{equation}
The subjectivity of this choice shows that age differences smaller than $\sim10^5\yr$ are not reliable in the present context,
and cannot be addressed by the present grid.
However, such delays are negligible from an observational point of view,
because on one hand the pre-MS time of low-mass stars is several orders of magnitude longer (typically above $10^7\yr$),
and on the other hand, the age spread between massive stars forming in a cluster is expected from hydrodynamical simulations
to be about $10^5\yr$ \citep{bonnell2003,peters2010a,peters2010b}.

\subsection{Masses}
\label{sec-m}

We consider the same masses as \cite{ekstroem2012}, which we write \Mac\ in the present work (see note \ref{not-m0}, Sect.~\ref{sec-dm}), namely
$\Mac=0.8$, 0.9, 1, 1.1, 1.25, 1.35, 1.5, 1.7, 2, 2.5, 3, 4, 5, 7, 9, 12, 15, 20, 25, 32, 40, 60, 85, 120 \Ms.
We emphasise that \Mac\ is the mass of the star at the end of the accretion phase.
These values cover most of the observed stellar mass range.

We also note that these masses correspond to the ZAMS mass \Mz\ only for the low- and intermediate-mass stars.
For the most massive stars, as mentioned in Sect.~\ref{sec-in}, the ZAMS is reached before the end of accretion,
and thus the mass continues to increase on the MS, so that $\Mac>\Mz$.

\subsection{Transition from pre-MS to MS}
\label{sec-zams}

A comparison between MS evolution starting from the ZAMS and that following the pre-MS phase
shows that the effect of pre-MS and accretion vanishes after a few percent of the MS lifetime.
This enables us to connect our pre-MS and early-MS models to the MS and previously published post-MS
grid \citep{ekstroem2012}.
However, since for the most massive models accretion overlaps the MS, we cannot make the connection on the ZAMS.
Instead, we connect the models at an arbitrary lower value $X_c=0.6$, after having checked that the models match.
Therefore, we emphasise that once $X_c<0.6$, the models of the present grid are strictly identical to those of \cite{ekstroem2012}.

\section{Grid}
\label{sec-grid}

\subsection{Birthline}
\label{sec-bl}

\begin{figure}
\includegraphics[width=0.49\textwidth]{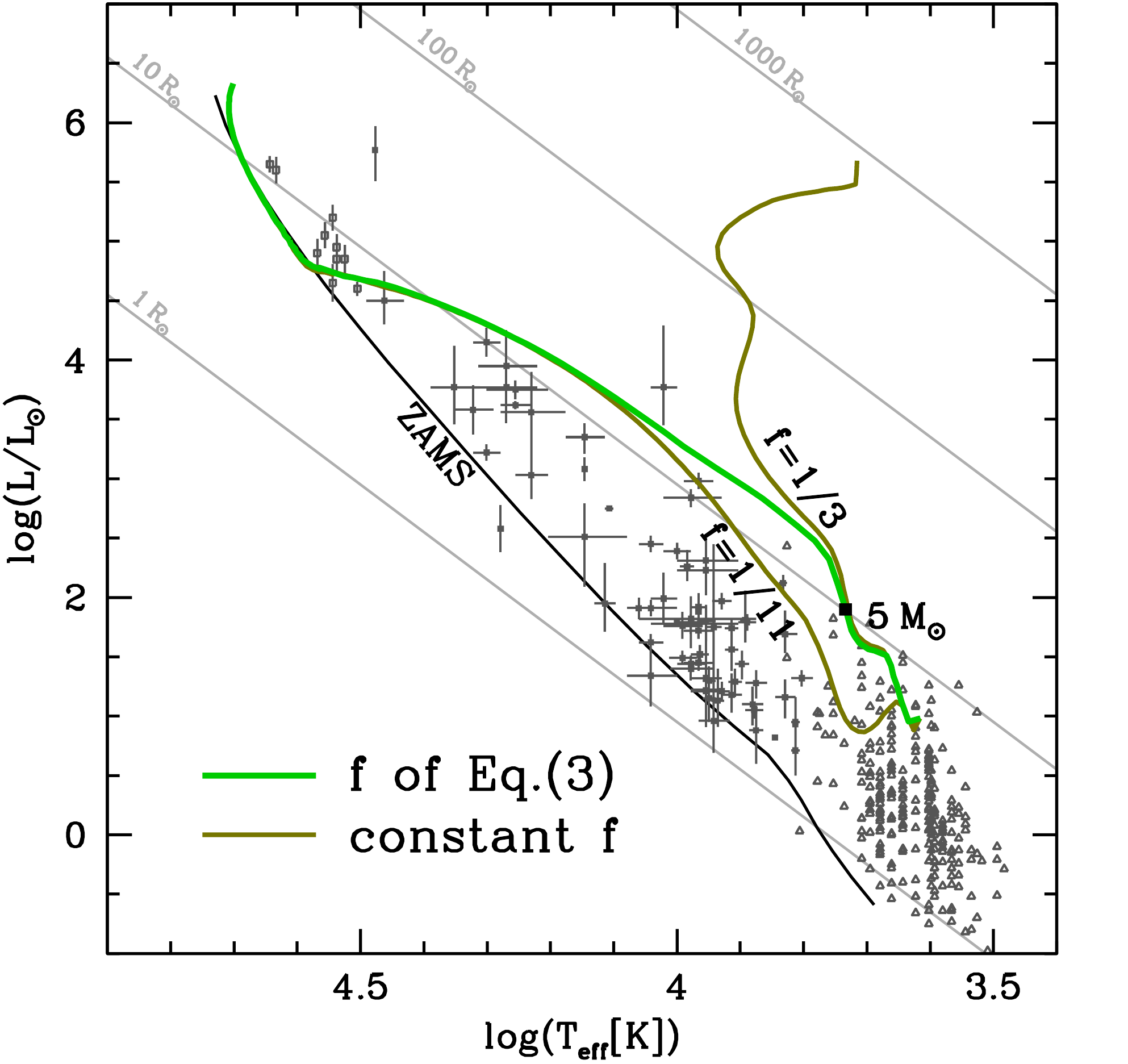}
\caption{Theoretical birthlines for various values of $f$.
The green track is the birthline used in the present grid, with $f$ given by Eq.~(\ref{eq-f}).
The brown tracks are birthlines with constant $f=1/3$ and $1/11$, from \cite{haemmerle2016a}.
The switching point of Eq.~(\ref{eq-f}) is indicated by a black square.
Other lines and points are identical to Fig.~\ref{fig-blobs}.}
\label{fig-blhr}\end{figure}

\begin{figure}
\includegraphics[width=0.49\textwidth]{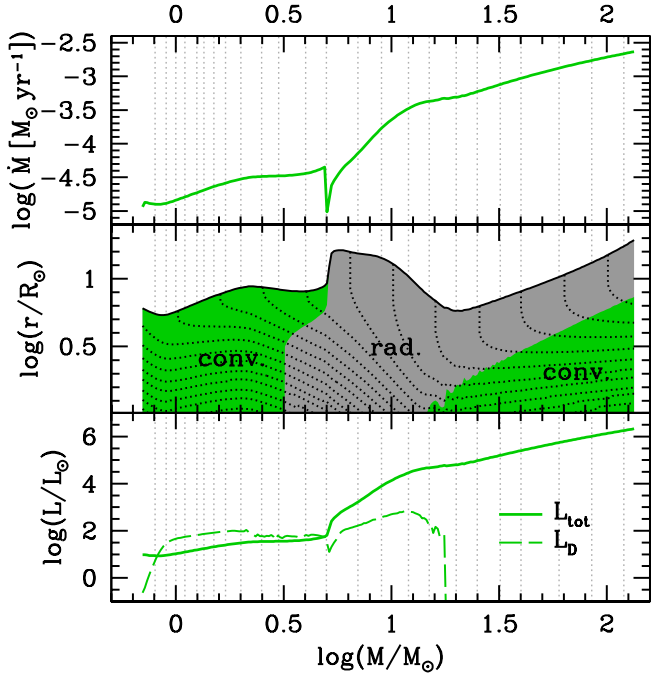}
\caption{Accretion rate, internal structure, and luminosity as a function of the mass on the birthline.
The vertical grey dotted straight lines indicate the masses \Mac\ of the model of the grid (Sect.~\ref{sec-m}).
On the middle panel, the upper black curve is the photospheric radius, the green (resp. grey) areas are convective (resp. radiative) regions,
and the black dotted curves are iso-mass (i.e. Lagrangian layers).
On the lower panel, the solid line is the total photospheric luminosity and the dashed one is the luminosity produced by D-burning.}
\label{fig-blst}\end{figure}

The evolutionary track and internal structure along the birthline are shown in Figs.~\ref{fig-blhr} and \ref{fig-blst}.
The properties of the birthline are similar to those described in \cite{haemmerle2016a}.
Here we recall the main features.

The initial structure is fully convective and the evolution starts by KH contraction.
While the star is in the low-mass range, it accretes at several $10^{-5}$~\Mpy.
D-burning starts rapidly in the centre, which causes the radius and luminosity to grow.
The luminosity from D-burning exceeds the total luminosity,
the difference being re-absorbed in the stellar interior as heat, and converted into gravitational energy through the expansion
(see the inner iso-masses on Fig.~\ref{fig-blst}).
After central D-exhaustion, all the layers shrink again and luminosity stops growing.
When the radiative core forms and grows in mass, the internal luminosity wave and shell D-burning cause rapid swelling
\citep{larson1972,hosokawa2009,hosokawa2010}.
Departure from thermal equilibrium is the strongest at this stage, and as a consequence
the radius and luminosity increase suddenly by a factor of two and by one order of magnitude, respectively.
The increase in luminosity induces an increase in the accretion rate, through the CH law (Eq.~\ref{eq-chout} and \ref{eq-chf}).
However, we switch from $f=1/3$ to $1/11$ at this point (Eq.~\ref{eq-f}), which limits this artificial loop.
When the convective envelope disappears, the surface contracts again,
but the luminosity and accretion rate continue to grow ($\dm\sim10^{-4}$~\Mpy).
During this phase, the star moves fast on the HR diagram, from $\logTeff=3.8$ to 4.6, and from $\log L/\Ls=2.5$ to 4.8.
The central temperature grows until H-burning starts in the centre and gives birth to a convective core.
The track has joined the ZAMS on the HR diagram, at a mass of about 20~\Ms.
After about $M=60$~\Ms, the birthline drifts slightly towards the right-hand side of the ZAMS,
partly due to MS evolution, partly to the high accretion rate ($\sim10^{-3}$~\Mpy)
which makes it hard for the surface to relax thermally on the non-accreting structure.
Indeed, in spite of the assumption of cold accretion,
the mass grows so fast that the internal entropy profile cannot converge on a non-accreting structure,
and thus the star keeps an excess of entropy.

In Fig.~\ref{fig-blhr}, we compare this birthline with the two birthlines at constant $f$ described in \cite{haemmerle2016a}.
In the beginning of the evolution the new birthline follows the one with $f=1/3$, according to Eq.~(\ref{eq-f}).
At $M=5$~\Ms, the value of $f$ switches to $1/11$ and the new birthline diverges from the $f=1/3$ one, with lower luminosities; it eventually joins the $f=1/11$ birthline at $\log L/\Ls\simeq3.5-4$, that is, at $M\simeq8$~\Ms, and follows it until the end of the run.
Figure~\ref{fig-blhr} also shows a set of observations of pre-MS stars (references given in the caption of Fig.~\ref{fig-blobs}).
Comparing these observations with the tracks,
we see that the new birthline approximately matches the upper envelope of the observations on the whole mass range.
In contrast, the $f=1/11$ birthline gives too low luminosities for $\logTeff<3.8$,
while the $f=1/3$ birthline diverges clearly from the observations at $\logTeff>3.8$.

\subsection{Evolutionary tracks}
\label{sec-trck}

\begin{figure*}\begin{center}
\includegraphics[width=0.8\textwidth]{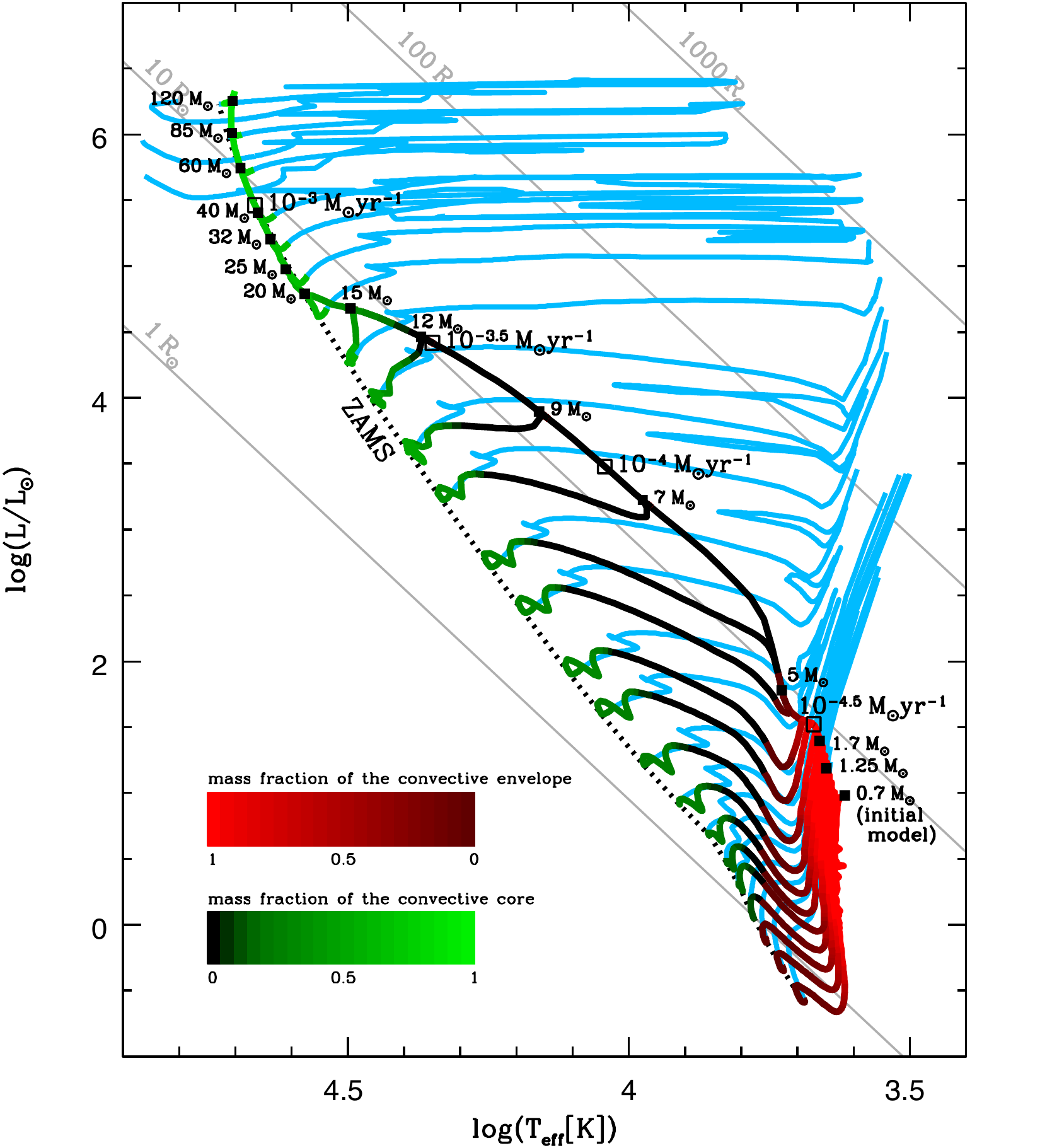}
\caption{HR diagram of the models.
The new tracks are plotted with a red-black-green colour gradient, with the indicated colour code:
the mass fraction of the convective envelope is indicated in red, while that of the convective core is coloured in green.
The black segments indicate fully radiative stars.
The mass and accretion rate along the birthline are indicated at various stages.
The blue curves are the non-rotating models of \cite{ekstroem2012}.
The black dotted line is the ZAMS of \cite{ekstroem2012}.
The grey straight lines are the iso-radii of the indicated radii.}
\label{fig-grid}\end{center}\end{figure*}

The evolutionary tracks of the models with various \Mac\ are shown in Fig.~\ref{fig-grid}.
The thick tracks in colour gradient (red-black-green) are the new models, that is, the pre-MS phase until a central mass fraction of hydrogen $X_c=0.6$,
the point at which we connect to the tracks of \cite{ekstroem2012}, which are plotted in blue.

Starting from the initial model (0.7\,\Ms), the evolutionary tracks of the various models follow the unique birthline (Sect.~\ref{sec-bl})
as accretion proceeds until $M=\Mac$.
When the various models reach their `final' masses \Mac, accretion stops and the track leaves the birthline to follow constant mass evolution,
or evolution with mass-loss.
The constant mass track depends on the value of \Mac:
\begin{itemize}
\item For $\Mac<3$~\Ms, the star is still fully convective when accretion stops (Fig.~\ref{fig-blst}), and the evolution switches along the Hayashi line,
with decreasing radii and luminosities and constant effective temperatures.
\item For $3~\Ms<\Mac<20~\Ms$, the star is mostly radiative (fully radiative for $5~\Ms<\Mac\leq12~\Ms$) at the end of accretion.
In this case, it evolves along the radiative branch with increasing effective temperatures and nearly constant luminosities.
\item For $\Mac>20$~\Ms, accretion stops after the track crossed the ZAMS in the HR diagram at $\logTeff\simeq4.5$.
Thus these models switch directly from accretion to MS evolution, without a constant-mass pre-MS phase.
\item For $\Mac\gtrsim70$~\Ms, central H depletion is already significant at the end of accretion ($X_c<0.716$; $X_c=0.713$ for 120~\Ms),
that is, for these models the accretion phase overlaps the early MS.
\item For the most massive model ($\Mac=120$~\Ms), the accretion rate is so high ($\simeq2\times10^{-3}$ \Mpy)
that the external layers inflate, because they radiate only a very small portion of the accreted entropy (Sect.~\ref{sec-bl}).
Once accretion stops, this entropy excess is radiated and the star converges to a non-accreting structure: it moves slightly to the left of the birthline.
After this short contraction however, the track follows classical MS evolution towards the red part of the HR diagram.
\end{itemize}

\subsection{Isochrones}
\label{sec-iso}

\begin{figure*}\begin{center}
\includegraphics[width=0.8\textwidth]{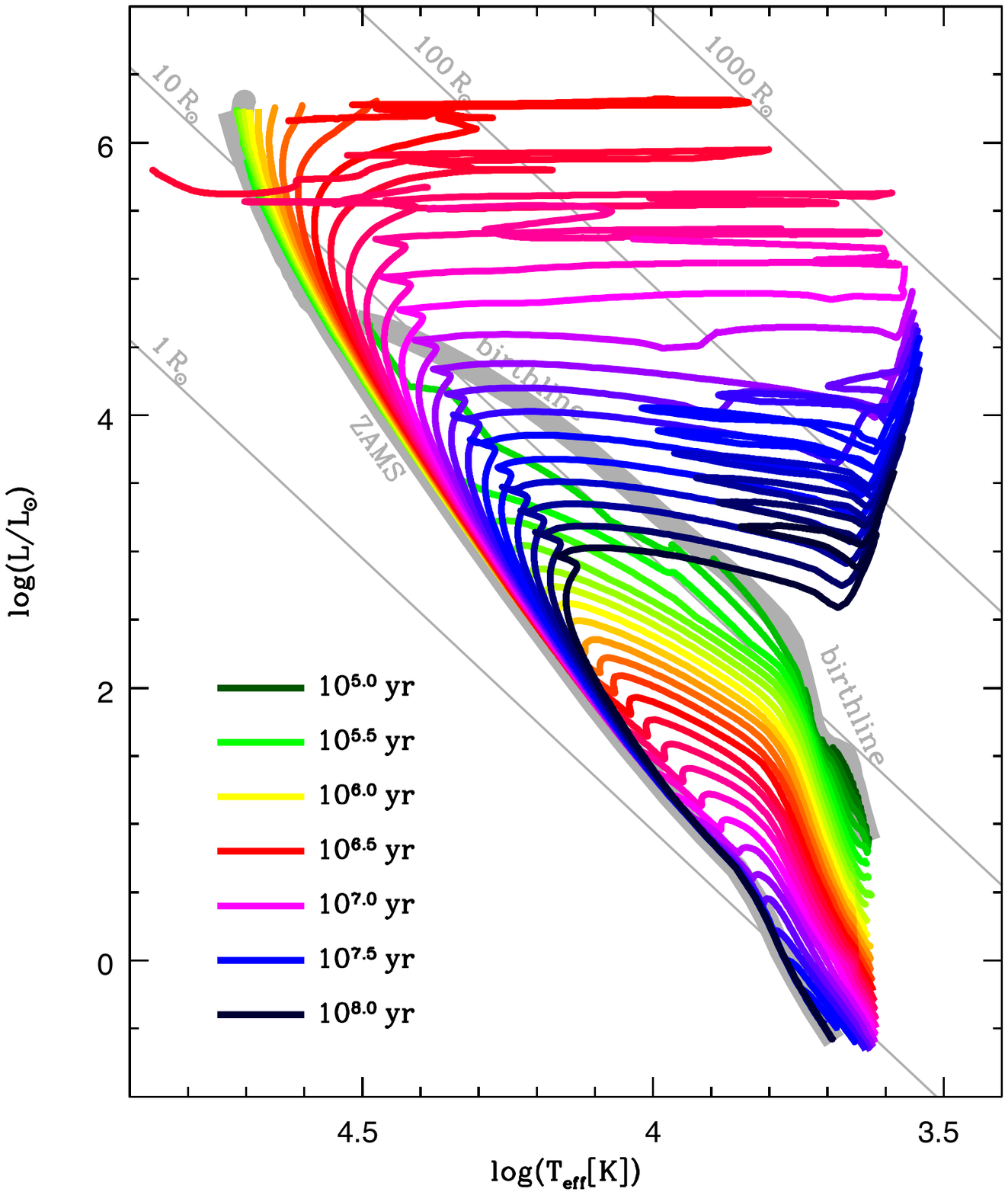}
\caption{Isochrones of $10^5\yr$ to $10^8\yr$, by steps of 0.1 dex.
The ZAMS of \cite{ekstroem2012} and the birthline of the present work are indicated by grey bands.
The grey straight lines are iso-radius of the indicated radii.}
\label{fig-iso}
\end{center}\end{figure*}

Figure~\ref{fig-iso} shows a series of isochrones derived from the grid (through \syc), for ages between $10^5$ and $10^8\yr$, by steps of 0.1 dex.
As explained in Sect.~\ref{sec-ini}, younger ages ($<10^5\yr$) are not relevant for the present grid
because of the uncertainty on early accretion history ($M<0.7$~\Ms).
For older ages ($>10^8\yr$), massive stars are dead, while low-mass stars already evolve on the MS.
The relevant isochrones for pre-MS are therefore in the range $10^5-10^8\yr$.
Here we describe the properties of a few isochrones.
\begin{itemize}
\item The isochrone $10^5\yr$ shows that at this young age all the models are still in the low-mass range ($M<1.5$~\Ms).
Due to the slow evolution for such masses, the isochrone still matches the lower tail of the birthline.
\item At $3\times10^5\yr$, all the models with $\Mac<20$~\Ms\ completed accretion.
This mass corresponds to almost the point where the birthline joins the ZAMS.
In the same time, the models with the lowest masses started contracting, but due to their long contraction time ($10^7-10^8\yr$),
they remain relatively close to the birthline.
Thus the isochrone $10^{5.5}\yr$  approximately follows the birthline from the low-mass end to the intersection with the ZAMS.
\item At $10^6\yr$, all the models have achieved accretion, and those with $\Mac\geq5$~\Ms\ have already reached the ZAMS.
The most massive model ($\Mac=120$~\Ms) spent about 25\% of its MS time, and has a central mass fraction of hydrogen of 0.6.
\item At $10^{6.5}\yr\simeq3.2\times10^6\yr$, the $\Mac=4$~\Ms\ model joined the ZAMS.
The $\Mac=120$~\Ms\ one exhausted its central hydrogen, and moved leftwards while burning helium.
In the same time, the $\Mac=85$~\Ms\ model approaches the end of MS, with a central mass fraction of hydrogen of 0.08,
and is thus located at the right end of the MS band.
\item At $10^7\yr$, all the models with $\Mac\geq20$~\Ms\ have reached the end of the evolution,
and all those with $2\,\Ms\leq\Mac<20$~\Ms\ are on the MS.
Only the low-mass models are still contracting towards the ZAMS.
\item In the last isochrone, at $10^8\yr$, all the models have passed the ZAMS, and all those with $\Mac>5$~\Ms\ have reached their terminal point.
The $\Mac=5$~\Ms\ model is burning helium in its core, while all lower-mass models are still on the MS.
\end{itemize}

\subsection{Online data}
\label{sec-tab}

Electronic tables of the models and isochrones are available online:
\href{https://www.unige.ch/sciences/astro/evolution/en/database/}{https://www.unige.ch/sciences/astro/evolution/en/database/}
These tables display the same quantities as in the MS and post-MS grids of \cite{ekstroem2012}, where each model is given by 400 data points.
Here we add the pre-MS phase through 100 new points.
Depending on \Mac, the first 10 -- 20 points of the MS phase must also be replaced using the new models,
since for the largest masses accretion continues on the MS (see Sect.~\ref{sec-zams}).

\section{Discussion}
\label{sec-discus}

\subsection{Impact of accretion on the evolutionary tracks and isochrones}
\label{sec-cst}

\begin{figure}
\includegraphics[width=0.49\textwidth]{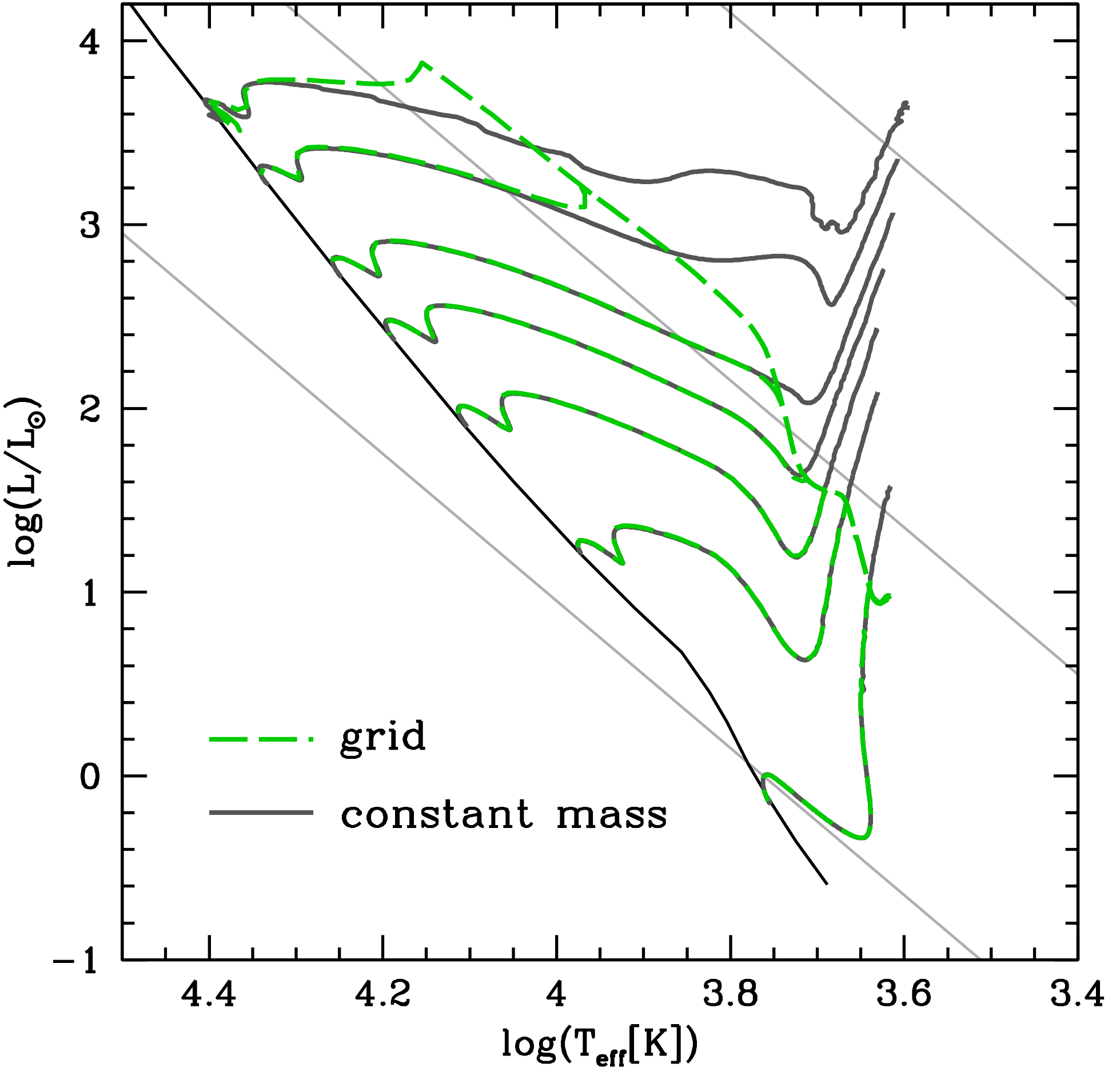}
\caption{Comparison between accreting and constant-mass evolutionary tracks on the pre-MS, for $\Mac=1$, 2, 3, 4, 5, 7 and 9 \Ms.}
\label{fig-hrcst}
\end{figure}

The effect of accretion on the pre-MS tracks is to cut the `upper' part of the tracks (Hayashi line, or radiative branch for massive stars),
because once accretion is completed, its centre has already evolved
\citep{larson1969,larson1972,bernasconi1996a,norberg2000,yorke2002,yorke2008}.
Figure~\ref{fig-hrcst} compares the evolutionary tracks of the present grid with constant mass tracks for $\Mac=1,$ 2, 3, 4, 5, 7 and 9~\Ms.
Models with higher masses are near the ZAMS when achieving accretion, and do not experience a significant pre-MS contraction at constant mass.

Since the KH time is shorter for larger masses, the massive models are more impacted by accretion than the low-mass ones.
For $\Mac\leq3$~\Ms, accretion stops while the star is still fully convective and constant-mass contraction starts by the (shortened) Hayashi track,
as described in Sect.~\ref{sec-trck}.
In this case, the location of the star at the end of accretion corresponds simply to the intersection between the birthline and the constant-mass track.
It shows that for low masses the structure of the star for given mass and entropy is not affected by the former accretion history.

For larger masses, the impact of accretion becomes visible in Fig.~\ref{fig-hrcst}.
These models do not have a Hayashi track, and switch directly from the birthline to the radiative branch.
For $\Mac>5$~\Ms\ accretion is completed only after the pre-MS swelling, when the rate grows above $\sim10^{-4}$~\Mpy\
and departure from thermal equilibrium is the strongest (Sect.~\ref{sec-bl}).
This means that the star hardly radiates the accreted entropy, and thus its structure reflects its previous accretion history, through an excess in luminosity.
We indeed see in Fig.~\ref{fig-hrcst} that, at the end of accretion, models with $\Mac=7$ and 9 \Ms\
are located slightly above their corresponding constant-mass track, indicating that the star is not thermally relaxed at this stage of the accretion phase (pre-MS swelling, see Sect.~\ref{sec-bl}),
that is, that there is an excess of entropy compared to non-accreting structures.
Subsequently, as the star contracts in a fraction of a KH time, the entropy excess is eventually radiated away,
and the track and structure relax on that of constant-mass evolution.
The structure on the ZAMS is unaffected by accretion.

\begin{figure}\begin{center}
\includegraphics[width=0.45\textwidth]{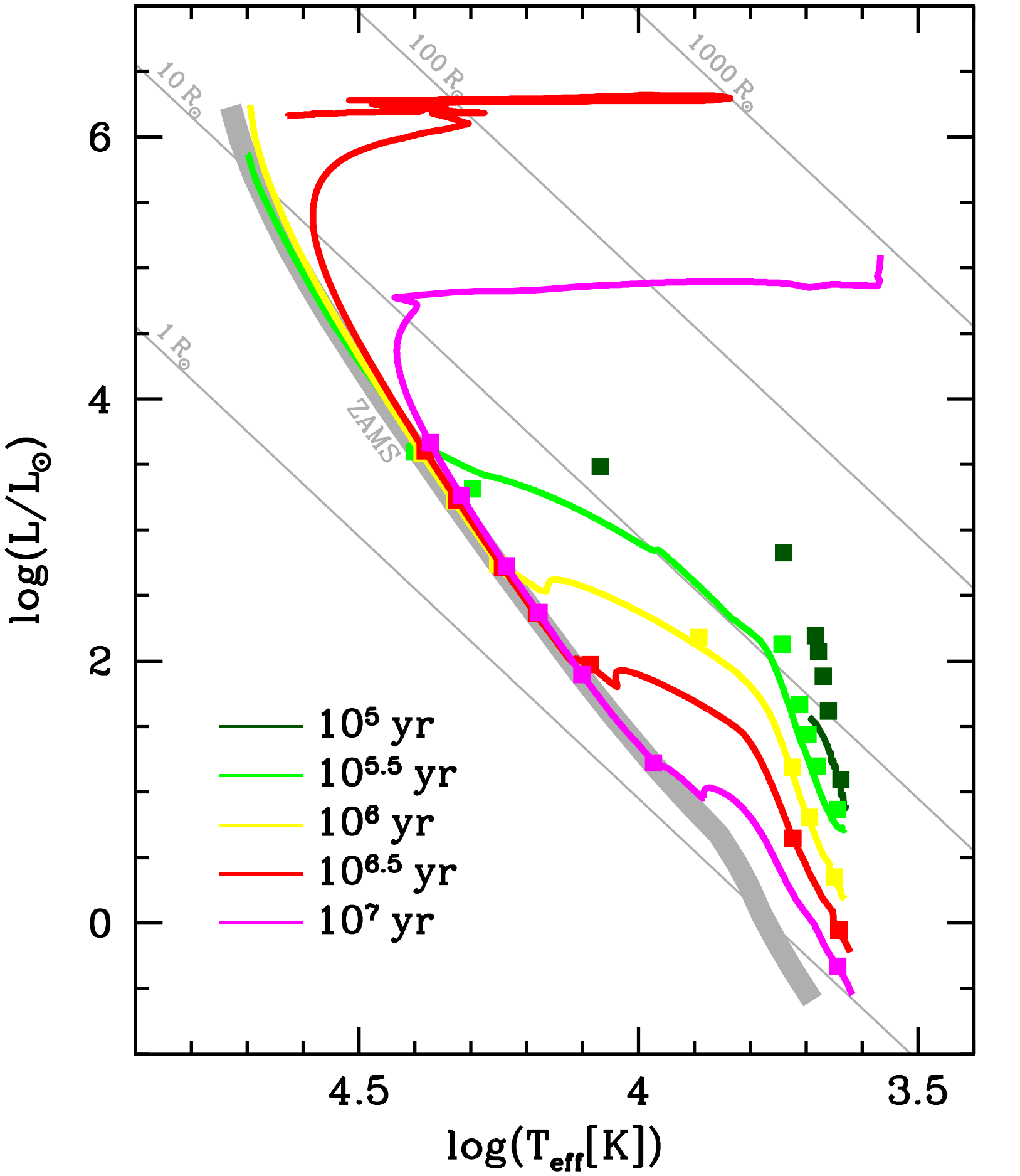}
\caption{Comparison between the isochrones of the grid (solid lines)
and isochrones without accretion ($\Mac=1,$ 2, 3, 4, 5, 7 and 9~\Ms, solid squares).}
\label{fig-isocst}
\end{center}\end{figure}

The impact of accretion on the isochrones is visible in Fig.~\ref{fig-isocst}.
The figure shows the isochrones of the grid (identical to Fig.~\ref{fig-iso}) for the indicated ages (solid lines)
and compares it with isochrones of the same age obtained with pre-MS evolution at constant mass,
for $\Mac=1,$ 2, 3, 4, 5, 7 and 9~\Ms.
This mass range is similar to that considered in the pre-MS grids available in the literature.
Obviously, since at ages of $\sim10^5$ years all the accreting models are still in the low-mass range,
the isochrone of $10^5$ yr extends to higher luminosities in the constant-mass case compared to the case with accretion.
However, already at 300 000 yr (light green), the differences between the isochrones at constant mass and those with accretion become negligible.
From $10^6$ yr onwards, the isochrones are identical.
Thus, the impact of accretion on the mass range $M\lesssim10$ \Ms\ is only significant  for very young ages ($\lesssim300\,000$ yr).
Compared to the grids available in the literature, the main advantage of the present grid
is to extend the isochrones to the range of massive stars ($M\gtrsim10$ \Ms) with the same stellar models.

\subsection{Impact of the accretion law}
\label{sec-ac2}

As explained in Sect.~\ref{sec-dm}, the CH relation does not necessarily correspond to an evolutionary sequence,
but might reflect the properties of stars with various masses during their main accretion phase.
In this alternative picture, stars destined to reach various masses by accretion do not follow the same accretion history.
Therefore, their accretion tracks are not unique and do not necessarily correspond  to the birthline.

In order to check the impact of the assumption of a unique accretion history,
we compute an alternative grid with the same physical inputs as the one described in the previous sections, except the accretion law.
In the new accretion law, the $M-\dm$ relation of Fig.~\ref{fig-dm} is applied with $M=\Mac$ instead of $M=M(t)$:
\begin{equation}
\dm(t)=\left\{\begin{array}{l}
\dm_{\rm CH}(M(t))\qquad{\rm(scenario\ 1)}\\
\dm_{\rm CH}(\Mac)\qquad\ \ \,{\rm(scenario\ 2)}\end{array}\right.\ .
\end{equation}
Thus, in scenario 2, stars destined to become massive start their accretion phase
with the high (constant) rates expected for massive star formation ($\sim10^{-3}$~\Mpy)
while in scenario 1 they first accrete at low rates ($\sim10^{-5}$~\Mpy) until they become effectively massive.
As a consequence, we expect massive stars with given $L$ and \Teff\ to be younger in scenario 2 than in scenario 1.
For this second grid, we consider masses $\Mac=1$, 5, 7, 12, 25 and 120~\Ms.

The change from scenario 1 to scenario 2 has no significant impact on the evolutionary tracks after accretion,
because, as discussed in Sect.~\ref{sec-cst}, the evolutionary tracks of models formed by accretion
converge rapidly to non-accreting tracks once accretion stops.

\begin{figure}
\includegraphics[width=0.49\textwidth]{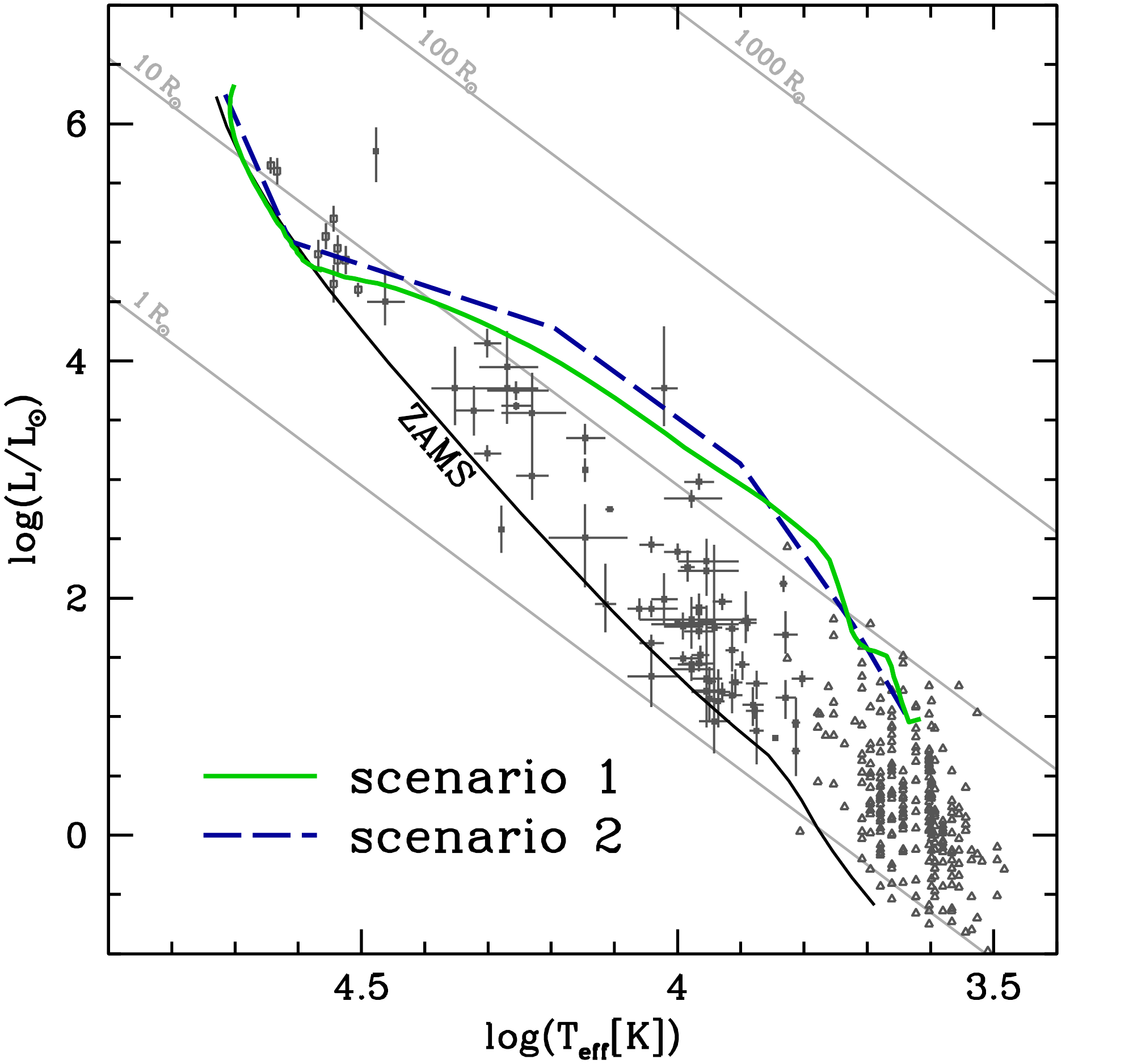}
\caption{Birthlines according to scenarios 1 and 2 (see Sect.~\ref{sec-ac2}).
The points are the same observations as in Fig.~\ref{fig-blobs} and \ref{fig-blhr}.}
\label{fig-bl2}
\end{figure}

Also, the birthline depends only weakly on the choice of the scenario.
Figure~\ref{fig-bl2} compares the birthlines obtained with both prescriptions.
In scenario 2, the birthline does not correspond to an accretion track, but sets the endpoints of the various accretion tracks.
We see that the birthline of scenario 2 does not differ significantly from that of scenario 1, and matches the observations as well.
In fact, the differences between the two birthlines are similar to the error bars of the observations.

The main difference between the two scenarios is expected to be age, in particular for the most massive models,
because their accretion history is more impacted by the change in the accretion law.
However, even taking this into consideration, the change in the accretion scenario has a limited impact.
This is because in scenario 1, the 120~\Ms\ model ends accretion at
\begin{equation}
t_1(120\,\Ms)=400\ 000\yr.
\end{equation}
In scenario 2, the age at which the 120~\Ms\ model ends accretion is necessarily shorter,
meaning that the shift between both models is at most a few $10^5\yr$.
Indeed, in scenario 2, the 120~\Ms\ forms at a constant rate of $2.17\times10^{-3}$~\Mpy.
Thus the time it takes to complete accretion is
\begin{equation}
t_2(120\,\Ms)={120\,\Ms\over2.2\times10^{-3}\,\Mpy}\simeq55\ 000\yr.
\label{eq-t2}\end{equation}
The shift in age between scenarios 1 and 2 is, for the most impacted model,  only about 345~000 yr.
As a consequence, the isochrones given by scenarios 1 and 2 differ only for ages of a few $10^5\yr$.

\begin{figure}
\includegraphics[width=0.49\textwidth]{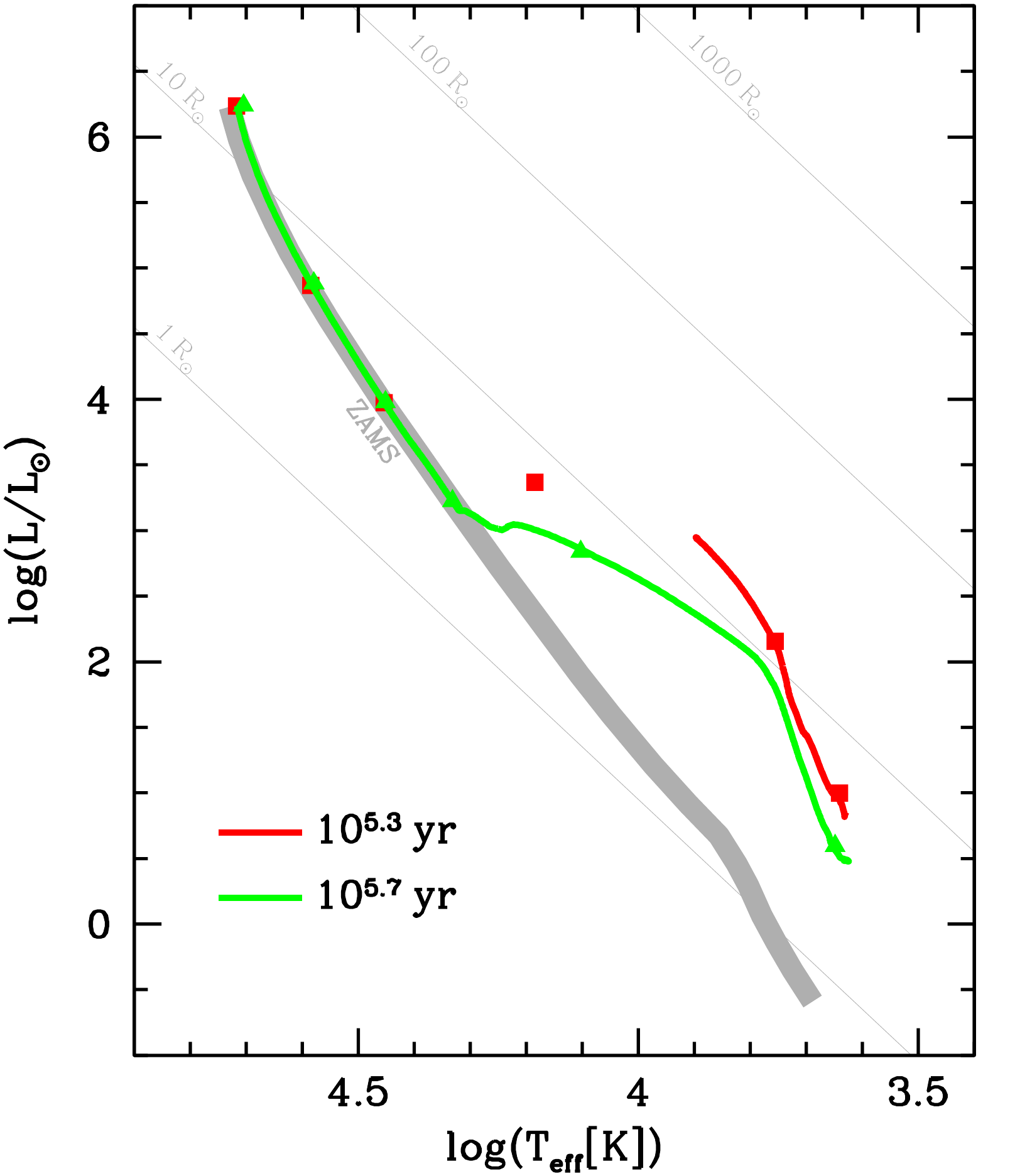}
\caption{Isochrones of $10^{5.3}\yr\simeq2\times10^5\yr$ and $10^{5.7}\yr\simeq5\times10^5\yr$ for the two scenarios described in Sect.~\ref{sec-ac2}.
Isochrones of scenario 1 are plotted with solid lines, while those of scenario 2 are shown by series of squares
($10^{5.3}\yr$) and triangles ($10^{5.7}\yr$).
The colours of the lines and squares indicate the age.}
\label{fig-iso2}
\end{figure}

Figure~\ref{fig-iso2} shows the isochrones of $2\times10^5\yr$ and $5\times10^5\yr$ in both scenarios.
Obviously, the isochrone $2\times10^5\yr$ reflects the change in the accretion law:
at this young age, in scenario 1, all the stars are still in the low-mass range (Sect.~\ref{sec-iso}),
while in scenario 2 the 120~\Ms\ model has already ended the accretion phase (Eq.~\ref{eq-t2}).
However, at $5\times10^5\yr$, the delay of massive models from scenario 1 is already negligible,
and isochrones of both grids coincide.
The reason is that at this age the 120~\Ms\ model is already on the MS, where it evolves according to its H-burning time, of about three million years, that is, one order of magnitude longer than the age shift between scenarios 1 and 2.

Interestingly, in scenario 2, all the models with different \Mac\ end accretion at the same time: about $10^5\yr$.
In scenario 1, massive stars end accretion after low-mass stars.
However, the rapid increase in the accretion rate as the mass grows (Fig.~\ref{fig-dm})
limits the delay of the optical appearance of massive stars with respect to low-mass stars.
This is the reason why the shift in age remains lower than a few times $10^5\yr$.
Beyond the particular choice of accretion scenario,
this increase in \dm\ reflects the presence of observed massive stars near the ZAMS at $\sim100$~\Ms\ (Fig.~\ref{fig-blobs}, \ref{fig-blhr} and \ref{fig-bl2}).
Any accretion law that matches these observations requires a high rate for massive star formation, that is, a short accretion timescale of about $10^5\yr$.
Such a timescale is similar to the age spread expected in massive star formation in a cluster (Sect.~\ref{sec-ini}).
Therefore, the choice of the accretion history is not able to impact the isochrones for ages above this limit,
as long as it reproduces the observations of massive young stellar objects.

\subsection{Isochrones: comparisons with observations}

\begin{figure*}
\includegraphics[width=0.49\textwidth]{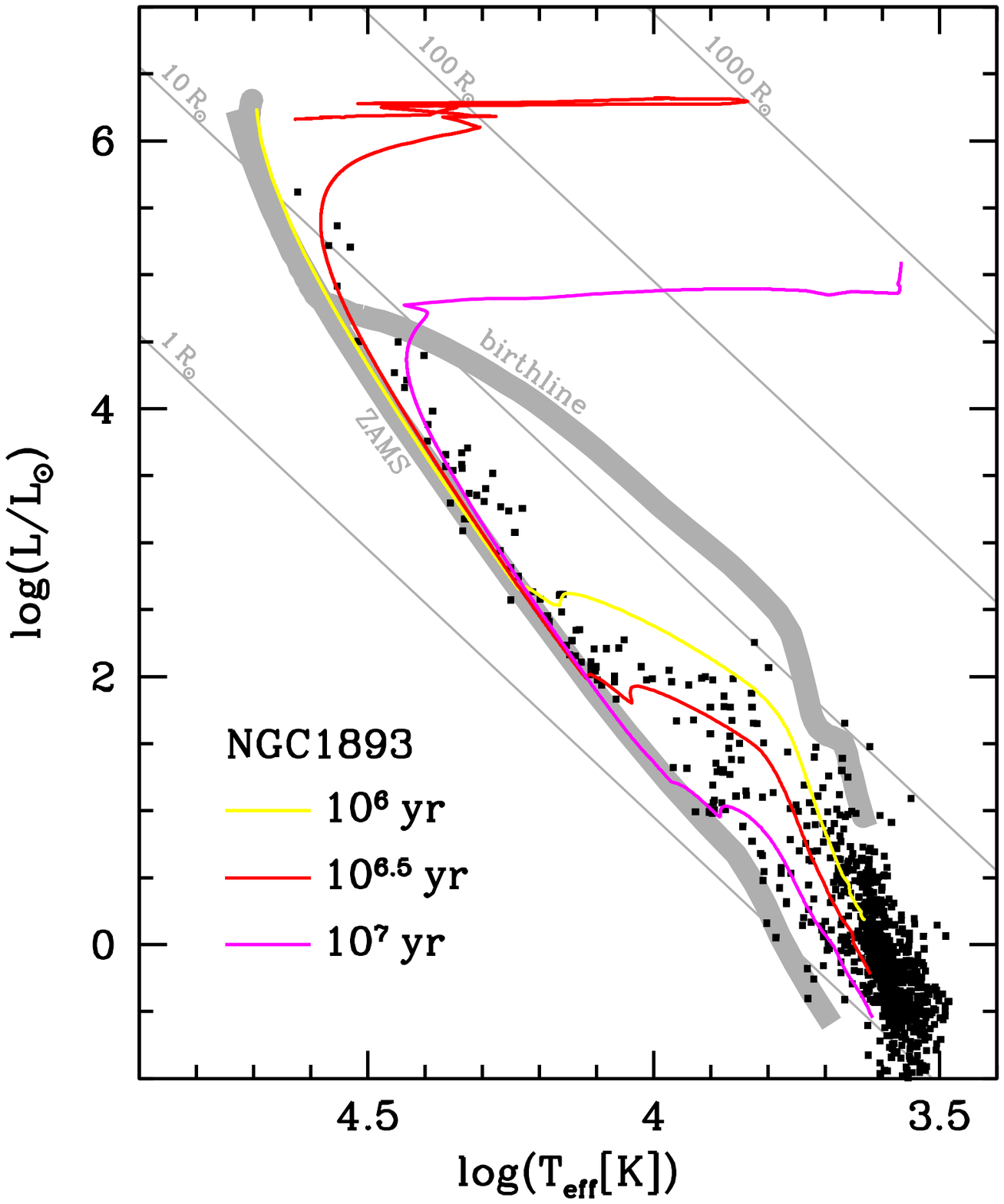}\hfill
\includegraphics[width=0.49\textwidth]{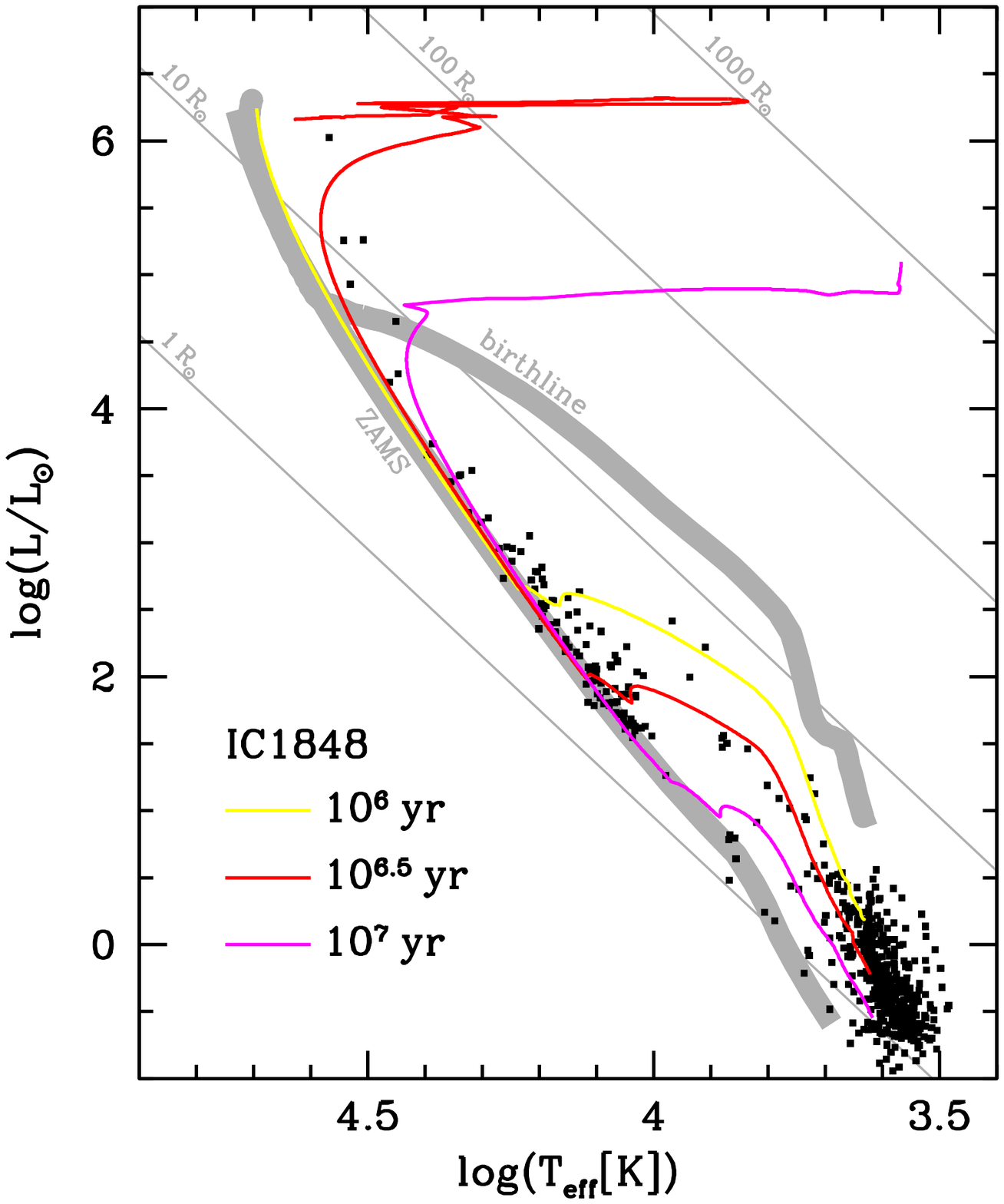}
\caption{Comparison between the isochrones of $10^6\yr$, $10^{6.5}\yr$ and $10^7\yr$ of the grid
and observations of the IC1848 and NGC1893 clusters, from \cite{lim2014a,lim2014b}.}
\label{fig-isoobs}
\end{figure*}

In order to test our method, we compared a set of isochrones with observations of young clusters, from the Sejong Open Cluster Survey
\citep{sung2013,lim2011,lim2014a,lim2014b,lim2015a,lim2015b}.
The comparison is shown in Fig.~\ref{fig-isoobs}.
We chose three isochrones with ages of $10^6\yr$, $10^{6.5}\yr,$ and $10^7\yr$.
As expected, the observations show a large dispersion, which could rely on a spread in age or in various other quantities.
But clearly, most of the points are located between the $10^6\yr$ and $10^7\yr$ isochrones, which indicates an age of several million years.
The $10^{6.5}\yr$ isochrone (i.e. $3\times10^6\yr$) appears to be a reasonable choice for both clusters.
However, for NGC1893, most of the points are located between the $10^6\yr$ and $10^{6.5}\yr$ isochrones,
meaning that we estimate the age to be about 2 Myr.
This is in agreement with the values 1.5 -- 1.9 Myr estimated by \cite{lim2014b} from the pre-MS isochrones at constant mass of \cite{siess2000}
and the MS and post-MS isochrones of \cite{ekstroem2012}.
For IC1848, more points are located between the $10^{6.5}\yr$ and $10^7\yr$ isochrones,
indicating an older age, about 3 -- 5 Myr, as estimated by \cite{lim2014a}.

\section{Summary and conclusions}
\label{sec-out}

We computed a pre-MS grid taking accretion into account,
over a wide mass-range from low-mass to massive stars ($0.8\,\Ms\leq M\leq120\,\Ms$) at solar metallicity.
This grid is connected with the MS and post-MS grid of \cite{ekstroem2012}, computed with the same physical inputs.
It provides a complete grid of stellar models covering most of the phases of stellar evolution,
from the early accretion to advanced stages (end of carbon-burning for massive stars, helium flash for low-mass stars).
Numerical tables of the models (including the birthline) and isochrones are available online
(\href{https://www.unige.ch/sciences/astro/evolution/en/database/}{https://www.unige.ch/sciences/astro/evolution/en/database/}
).

Accretion has been included through the CH accretion law, tuned to match the observational birthline on the HR diagram.
This law gives an accretion rate that increases with the mass of the accretor, from $\sim10^{-5}$~\Mpy\ for low-mass stars
to $10^{-3}$~\Mpy\ for massive stars, in agreement with the usual picture of star formation.
We assumed a unique accretion history, independent of the total mass the star is destined to accrete.
We checked how relaxing this assumption impacts the tracks and isochrones, and found the effect not to be significant.
Moreover, we obtained that the effect of accretion on stellar structure is significant only during the accretion phase,
and that once accretion is achieved, the tracks converge rapidly to the canonical tracks and isochrones of constant-mass evolution on the pre-MS.

Finally, we used our isochrones to estimate the age of the young clusters NGC1893 and IC 1848 from observations by \cite{lim2014a,lim2014b}.
We obtained ages of $\sim2$ and $\sim4$ Myr, in agreement with previous estimates.

Rotation is not included in the present grid.
We postpone to a forthcoming work the description of a grid with the same masses and metallicity as the present one but
with rotation in addition.

\begin{acknowledgements}
This work was sponsored by the Swiss National Science Foundation (project number 200020-172505).
AP and MA acknowledge support from the Swiss National Science Foundation (SNSF) (project number 200021L\_163172).
We warmly thank Dr. Beomdu Lim and Prof. Hwankyung Sung for providing the observational data used in Fig.~\ref{fig-isoobs}.
\end{acknowledgements}

\bibliographystyle{aa}
\bibliography{bibliotheque}

\begin{thebibliography}{70}
\expandafter\ifx\csname natexlab\endcsname\relax\def\natexlab#1{#1}\fi

\bibitem[{{Alecian} {et~al.}(2013){Alecian}, {Wade}, {Catala}, {Grunhut},
  {Landstreet}, {Bagnulo}, {B{\"o}hm}, {Folsom}, {Marsden}, \&
  {Waite}}]{alecian2013a}
{Alecian}, E., {Wade}, G.~A., {Catala}, C., {et~al.} 2013, \mnras, 429, 1001

\bibitem[{{Asplund} {et~al.}(2005){Asplund}, {Grevesse}, \&
  {Sauval}}]{asplund2005}
{Asplund}, M., {Grevesse}, N., \& {Sauval}, A.~J. 2005, in Astronomical Society
  of the Pacific Conference Series, Vol. 336, Cosmic Abundances as Records of
  Stellar Evolution and Nucleosynthesis, ed. T.~G. {Barnes}, III \& F.~N.
  {Bash}, 25

\bibitem[{{Baraffe} {et~al.}(2009){Baraffe}, {Chabrier}, \&
  {Gallardo}}]{baraffe2009}
{Baraffe}, I., {Chabrier}, G., \& {Gallardo}, J. 2009, \apjl, 702, L27

\bibitem[{{Baraffe} {et~al.}(2012){Baraffe}, {Vorobyov}, \&
  {Chabrier}}]{baraffe2012}
{Baraffe}, I., {Vorobyov}, E., \& {Chabrier}, G. 2012, \apj, 756, 118

\bibitem[{{Behrend} \& {Maeder}(2001)}]{behrend2001}
{Behrend}, R. \& {Maeder}, A. 2001, \aap, 373, 190

\bibitem[{{Bernasconi}(1996)}]{bernasconi1996b}
{Bernasconi}, P.~A. 1996, \aaps, 120, 57

\bibitem[{{Bernasconi} \& {Maeder}(1996)}]{bernasconi1996a}
{Bernasconi}, P.~A. \& {Maeder}, A. 1996, \aap, 307, 829

\bibitem[{{Bondi}(1952)}]{bondi1952}
{Bondi}, H. 1952, \mnras, 112, 195

\bibitem[{{Bondi} \& {Hoyle}(1944)}]{bondi1944}
{Bondi}, H. \& {Hoyle}, F. 1944, \mnras, 104, 273

\bibitem[{{Bonnell} {et~al.}(2003){Bonnell}, {Bate}, \& {Vine}}]{bonnell2003}
{Bonnell}, I.~A., {Bate}, M.~R., \& {Vine}, S.~G. 2003, \mnras, 343, 413

\bibitem[{{Caughlan} \& {Fowler}(1988)}]{caughlan1988}
{Caughlan}, G.~R. \& {Fowler}, W.~A. 1988, Atomic Data and Nuclear Data Tables,
  40, 283

\bibitem[{{Churchwell}(1999)}]{churchwell1999}
{Churchwell}, E. 1999, in NATO ASIC Proc. 540: The Origin of Stars and
  Planetary Systems, ed. C.~J. {Lada} \& N.~D. {Kylafis}, 515

\bibitem[{{Cohen} \& {Kuhi}(1979)}]{cohen1979}
{Cohen}, M. \& {Kuhi}, L.~V. 1979, \apjs, 41, 743

\bibitem[{{Cunha} {et~al.}(2006){Cunha}, {Hubeny}, \& {Lanz}}]{cunha2006}
{Cunha}, K., {Hubeny}, I., \& {Lanz}, T. 2006, \apjl, 647, L143

\bibitem[{{Davies} {et~al.}(2011){Davies}, {Hoare}, {Lumsden}, {Hosokawa},
  {Oudmaijer}, {Urquhart}, {Mottram}, \& {Stead}}]{davies2011}
{Davies}, B., {Hoare}, M.~G., {Lumsden}, S.~L., {et~al.} 2011, \mnras, 416, 972

\bibitem[{{Eggenberger} {et~al.}(2008){Eggenberger}, {Meynet}, {Maeder},
  {Hirschi}, {Charbonnel}, {Talon}, \& {Ekstr{\"o}m}}]{eggenberger2008}
{Eggenberger}, P., {Meynet}, G., {Maeder}, A., {et~al.} 2008, \apss, 316, 43

\bibitem[{{Ekstr{\"o}m} {et~al.}(2012){Ekstr{\"o}m}, {Georgy}, {Eggenberger},
  {Meynet}, {Mowlavi}, {Wyttenbach}, {Granada}, {Decressin}, {Hirschi},
  {Frischknecht}, {Charbonnel}, \& {Maeder}}]{ekstroem2012}
{Ekstr{\"o}m}, S., {Georgy}, C., {Eggenberger}, P., {et~al.} 2012, \aap, 537,
  A146

\bibitem[{{Georgy} {et~al.}(2013{\natexlab{a}}){Georgy}, {Ekstr{\"o}m},
  {Eggenberger}, {Meynet}, {Haemmerl{\'e}}, {Maeder}, {Granada}, {Groh},
  {Hirschi}, {Mowlavi}, {Yusof}, {Charbonnel}, {Decressin}, \&
  {Barblan}}]{georgy2013b}
{Georgy}, C., {Ekstr{\"o}m}, S., {Eggenberger}, P., {et~al.}
  2013{\natexlab{a}}, \aap, 558, A103

\bibitem[{{Georgy} {et~al.}(2013{\natexlab{b}}){Georgy}, {Ekstr{\"o}m},
  {Granada}, {Meynet}, {Mowlavi}, {Eggenberger}, \& {Maeder}}]{georgy2013a}
{Georgy}, C., {Ekstr{\"o}m}, S., {Granada}, A., {et~al.} 2013{\natexlab{b}},
  \aap, 553, A24

\bibitem[{{Georgy} {et~al.}(2012){Georgy}, {Ekstr{\"o}m}, {Meynet}, {Massey},
  {Levesque}, {Hirschi}, {Eggenberger}, \& {Maeder}}]{georgy2012}
{Georgy}, C., {Ekstr{\"o}m}, S., {Meynet}, G., {et~al.} 2012, \aap, 542, A29

\bibitem[{{Girichidis} {et~al.}(2012{\natexlab{a}}){Girichidis}, {Federrath},
  {Allison}, {Banerjee}, \& {Klessen}}]{girichidis2012b}
{Girichidis}, P., {Federrath}, C., {Allison}, R., {Banerjee}, R., \& {Klessen},
  R.~S. 2012{\natexlab{a}}, \mnras, 420, 3264

\bibitem[{{Girichidis} {et~al.}(2011){Girichidis}, {Federrath}, {Banerjee}, \&
  {Klessen}}]{girichidis2011}
{Girichidis}, P., {Federrath}, C., {Banerjee}, R., \& {Klessen}, R.~S. 2011,
  \mnras, 413, 2741

\bibitem[{{Girichidis} {et~al.}(2012{\natexlab{b}}){Girichidis}, {Federrath},
  {Banerjee}, \& {Klessen}}]{girichidis2012a}
{Girichidis}, P., {Federrath}, C., {Banerjee}, R., \& {Klessen}, R.~S.
  2012{\natexlab{b}}, \mnras, 420, 613

\bibitem[{{Granada} \& {Haemmerl{\'e}}(2014)}]{granada2014}
{Granada}, A. \& {Haemmerl{\'e}}, L. 2014, \aap, 570, A18

\bibitem[{{Haemmerl{\'e}}(2014)}]{haemmerle2014}
{Haemmerl{\'e}}, L. 2014, PhD thesis, University of Geneva

\bibitem[{{Haemmerl{\'e}} {et~al.}(2013){Haemmerl{\'e}}, {Eggenberger},
  {Meynet}, {Maeder}, \& {Charbonnel}}]{haemmerle2013}
{Haemmerl{\'e}}, L., {Eggenberger}, P., {Meynet}, G., {Maeder}, A., \&
  {Charbonnel}, C. 2013, \aap, 557, A112

\bibitem[{{Haemmerl{\'e}} {et~al.}(2016){Haemmerl{\'e}}, {Eggenberger},
  {Meynet}, {Maeder}, \& {Charbonnel}}]{haemmerle2016a}
{Haemmerl{\'e}}, L., {Eggenberger}, P., {Meynet}, G., {Maeder}, A., \&
  {Charbonnel}, C. 2016, \aap, 585, A65

\bibitem[{{Haemmerl{\'e}} {et~al.}(2017){Haemmerl{\'e}}, {Eggenberger},
  {Meynet}, {Maeder}, {Charbonnel}, \& {Klessen}}]{haemmerle2017}
{Haemmerl{\'e}}, L., {Eggenberger}, P., {Meynet}, G., {et~al.} 2017, \aap, 602,
  A17

\bibitem[{{Haemmerl{\'e}} \& {Peters}(2016)}]{haemmerle2016b}
{Haemmerl{\'e}}, L. \& {Peters}, T. 2016, \mnras, 458, 3299

\bibitem[{{Hayashi}(1961)}]{hayashi1961b}
{Hayashi}, C. 1961, \pasj, 13, 450

\bibitem[{{Henning} {et~al.}(2000){Henning}, {Schreyer}, {Launhardt}, \&
  {Burkert}}]{henning2000}
{Henning}, T., {Schreyer}, K., {Launhardt}, R., \& {Burkert}, A. 2000, \aap,
  353, 211

\bibitem[{{Hosokawa} {et~al.}(2011){Hosokawa}, {Offner}, \&
  {Krumholz}}]{hosokawa2011a}
{Hosokawa}, T., {Offner}, S.~S.~R., \& {Krumholz}, M.~R. 2011, \apj, 738, 140

\bibitem[{{Hosokawa} \& {Omukai}(2009)}]{hosokawa2009}
{Hosokawa}, T. \& {Omukai}, K. 2009, \apj, 691, 823

\bibitem[{{Hosokawa} {et~al.}(2010){Hosokawa}, {Yorke}, \&
  {Omukai}}]{hosokawa2010}
{Hosokawa}, T., {Yorke}, H.~W., \& {Omukai}, K. 2010, \apj, 721, 478

\bibitem[{{Krumholz} {et~al.}(2007){Krumholz}, {Klein}, \&
  {McKee}}]{krumholz2007}
{Krumholz}, M.~R., {Klein}, R.~I., \& {McKee}, C.~F. 2007, \apj, 656, 959

\bibitem[{{Krumholz} {et~al.}(2009){Krumholz}, {Klein}, {McKee}, {Offner}, \&
  {Cunningham}}]{krumholz2009}
{Krumholz}, M.~R., {Klein}, R.~I., {McKee}, C.~F., {Offner}, S.~S.~R., \&
  {Cunningham}, A.~J. 2009, Science, 323, 754

\bibitem[{{Kuiper} {et~al.}(2010){Kuiper}, {Klahr}, {Beuther}, \&
  {Henning}}]{kuiper2010b}
{Kuiper}, R., {Klahr}, H., {Beuther}, H., \& {Henning}, T. 2010, \apj, 722,
  1556

\bibitem[{{Kuiper} {et~al.}(2011){Kuiper}, {Klahr}, {Beuther}, \&
  {Henning}}]{kuiper2011}
{Kuiper}, R., {Klahr}, H., {Beuther}, H., \& {Henning}, T. 2011, \apj, 732, 20

\bibitem[{{Larson}(1969)}]{larson1969}
{Larson}, R.~B. 1969, \mnras, 145, 271

\bibitem[{{Larson}(1972)}]{larson1972}
{Larson}, R.~B. 1972, \mnras, 157, 121

\bibitem[{{Lim} {et~al.}(2015{\natexlab{a}}){Lim}, {Sung}, {Bessell}, {Kim},
  {Hur}, \& {Park}}]{lim2015a}
{Lim}, B., {Sung}, H., {Bessell}, M.~S., {et~al.} 2015{\natexlab{a}}, \aj, 149,
  127

\bibitem[{{Lim} {et~al.}(2015{\natexlab{b}}){Lim}, {Sung}, {Hur}, {Lee},
  {Bessell}, {Kim}, {Lee}, {Park}, \& {Jeong}}]{lim2015b}
{Lim}, B., {Sung}, H., {Hur}, H., {et~al.} 2015{\natexlab{b}}, Journal of
  Korean Astronomical Society, 48, 343

\bibitem[{{Lim} {et~al.}(2014{\natexlab{a}}){Lim}, {Sung}, {Kim}, {Bessell}, \&
  {Karimov}}]{lim2014a}
{Lim}, B., {Sung}, H., {Kim}, J.~S., {Bessell}, M.~S., \& {Karimov}, R.
  2014{\natexlab{a}}, \mnras, 438, 1451

\bibitem[{{Lim} {et~al.}(2014{\natexlab{b}}){Lim}, {Sung}, {Kim}, {Bessell}, \&
  {Park}}]{lim2014b}
{Lim}, B., {Sung}, H., {Kim}, J.~S., {Bessell}, M.~S., \& {Park}, B.-G.
  2014{\natexlab{b}}, \mnras, 443, 454

\bibitem[{{Lim} {et~al.}(2011){Lim}, {Sung}, {Karimov}, \&
  {Ibrahimov}}]{lim2011}
{Lim}, B., {Sung}, H.~S., {Karimov}, R., \& {Ibrahimov}, M. 2011, Journal of
  Korean Astronomical Society, 44, 39

\bibitem[{{Martins} {et~al.}(2012){Martins}, {Mahy}, {Hillier}, \&
  {Rauw}}]{martins2012}
{Martins}, F., {Mahy}, L., {Hillier}, D.~J., \& {Rauw}, G. 2012, \aap, 538, A39

\bibitem[{{McKee} \& {Tan}(2003)}]{mckee2003}
{McKee}, C.~F. \& {Tan}, J.~C. 2003, \apj, 585, 850

\bibitem[{{Meyer} {et~al.}(2018){Meyer}, {Kuiper}, {Kley}, {Johnston}, \&
  {Vorobyov}}]{meyer2018}
{Meyer}, D.~M.-A., {Kuiper}, R., {Kley}, W., {Johnston}, K.~G., \& {Vorobyov},
  E. 2018, \mnras, 473, 3615

\bibitem[{{Meyer} {et~al.}(2017){Meyer}, {Vorobyov}, {Kuiper}, \&
  {Kley}}]{meyer2017}
{Meyer}, D.~M.-A., {Vorobyov}, E.~I., {Kuiper}, R., \& {Kley}, W. 2017, \mnras,
  464, L90

\bibitem[{{Mottram} {et~al.}(2010){Mottram}, {Hoare}, {Lumsden}, {Oudmaijer},
  {Urquhart}, {Meade}, {Moore}, \& {Stead}}]{mottram2010}
{Mottram}, J.~C., {Hoare}, M.~G., {Lumsden}, S.~L., {et~al.} 2010, \aap, 510,
  A89

\bibitem[{{Mowlavi} {et~al.}(2012){Mowlavi}, {Eggenberger}, {Meynet},
  {Ekstr{\"o}m}, {Georgy}, {Maeder}, {Charbonnel}, \& {Eyer}}]{mowlavi2012}
{Mowlavi}, N., {Eggenberger}, P., {Meynet}, G., {et~al.} 2012, \aap, 541, A41

\bibitem[{{Norberg} \& {Maeder}(2000)}]{norberg2000}
{Norberg}, P. \& {Maeder}, A. 2000, \aap, 359, 1025

\bibitem[{{Palla} \& {Stahler}(1990)}]{palla1990}
{Palla}, F. \& {Stahler}, S.~W. 1990, \apjl, 360, L47

\bibitem[{{Palla} \& {Stahler}(1992)}]{palla1992}
{Palla}, F. \& {Stahler}, S.~W. 1992, \apj, 392, 667

\bibitem[{{Peters} {et~al.}(2011){Peters}, {Banerjee}, {Klessen}, \& {Mac
  Low}}]{peters2011}
{Peters}, T., {Banerjee}, R., {Klessen}, R.~S., \& {Mac Low}, M.-M. 2011, \apj,
  729, 72

\bibitem[{{Peters} {et~al.}(2010{\natexlab{a}}){Peters}, {Banerjee}, {Klessen},
  {Mac Low}, {Galv{\'a}n-Madrid}, \& {Keto}}]{peters2010a}
{Peters}, T., {Banerjee}, R., {Klessen}, R.~S., {et~al.} 2010{\natexlab{a}},
  \apj, 711, 1017

\bibitem[{{Peters} {et~al.}(2010{\natexlab{b}}){Peters}, {Klessen}, {Mac Low},
  \& {Banerjee}}]{peters2010c}
{Peters}, T., {Klessen}, R.~S., {Mac Low}, M.-M., \& {Banerjee}, R.
  2010{\natexlab{b}}, \apj, 725, 134

\bibitem[{{Peters} {et~al.}(2010{\natexlab{c}}){Peters}, {Mac Low}, {Banerjee},
  {Klessen}, \& {Dullemond}}]{peters2010b}
{Peters}, T., {Mac Low}, M.-M., {Banerjee}, R., {Klessen}, R.~S., \&
  {Dullemond}, C.~P. 2010{\natexlab{c}}, \apj, 719, 831

\bibitem[{{Schmeja} \& {Klessen}(2004)}]{schmeja2004}
{Schmeja}, S. \& {Klessen}, R.~S. 2004, \aap, 419, 405

\bibitem[{{Siess} {et~al.}(2000){Siess}, {Dufour}, \& {Forestini}}]{siess2000}
{Siess}, L., {Dufour}, E., \& {Forestini}, M. 2000, \aap, 358, 593

\bibitem[{{Siess} {et~al.}(1997){Siess}, {Forestini}, \& {Bertout}}]{siess1997}
{Siess}, L., {Forestini}, M., \& {Bertout}, C. 1997, \aap, 326, 1001

\bibitem[{{S{\o}rensen} {et~al.}(2018){S{\o}rensen}, {Fragos}, {Meynet}, \&
  {Haemmerl{\'e}}}]{sorensen2018}
{S{\o}rensen}, M., {Fragos}, T., {Meynet}, G., \& {Haemmerl{\'e}}, L. 2018,
  arXiv e-prints

\bibitem[{{Stahler}(1983)}]{stahler1983}
{Stahler}, S.~W. 1983, \apj, 274, 822

\bibitem[{{Stahler}(1988)}]{stahler1988}
{Stahler}, S.~W. 1988, \apj, 332, 804

\bibitem[{{Sung} {et~al.}(2013){Sung}, {Lim}, {Bessell}, {Kim}, {Hur}, {Chun},
  \& {Park}}]{sung2013}
{Sung}, H., {Lim}, B., {Bessell}, M.~S., {et~al.} 2013, Journal of Korean
  Astronomical Society, 46, 103

\bibitem[{{Tognelli} {et~al.}(2015){Tognelli}, {Prada Moroni}, \&
  {Degl'Innocenti}}]{tognelli2015}
{Tognelli}, E., {Prada Moroni}, P.~G., \& {Degl'Innocenti}, S. 2015, \mnras,
  454, 4037

\bibitem[{{Yang} {et~al.}(2018){Yang}, {Thompson}, {Urquhart}, \&
  {Tian}}]{yang2017}
{Yang}, A.~Y., {Thompson}, M.~A., {Urquhart}, J.~S., \& {Tian}, W.~W. 2018,
  \apjs, 235, 3

\bibitem[{{Yorke} \& {Bodenheimer}(2008)}]{yorke2008}
{Yorke}, H.~W. \& {Bodenheimer}, P. 2008, in Astronomical Society of the
  Pacific Conference Series, Vol. 387, Massive Star Formation: Observations
  Confront Theory, ed. H.~{Beuther}, H.~{Linz}, \& T.~{Henning}, 189

\bibitem[{{Yorke} \& {Kruegel}(1977)}]{yorke1977}
{Yorke}, H.~W. \& {Kruegel}, E. 1977, \aap, 54, 183

\bibitem[{{Yorke} \& {Sonnhalter}(2002)}]{yorke2002}
{Yorke}, H.~W. \& {Sonnhalter}, C. 2002, \apj, 569, 846

\end{thebibliography}

\end{document}